\documentclass[a4paper]{article}
\pdfoutput=1
\usepackage[utf8x]{inputenc}
\usepackage{amssymb,amsfonts,amsmath,amstext,amsgen,amsopn,amsxtra,indentfirst,graphicx,color,ulem}
\usepackage{xspace}
\usepackage{ulem}
\usepackage{hyperref}
\usepackage{graphicx,subcaption,float}
\usepackage{geometry}
\usepackage[numbers,super,comma, compress]{natbib}

\usepackage{comment,hyperref,url}
\newcommand{\kmpers}{\,km\,s$^{-1}$\xspace}
\newcommand{\vorb}{$200.7 \pm 4.9$\xspace}
\newcommand{\vorbprecise}{$194.1 \pm 4.3$\xspace}
\newcommand{\aj}{Astron. J.}   
\newcommand{\apj}{Astrophys. J.}   
\newcommand{\apjl}{Astrophys. J. Lett.}   
\newcommand{\apjs}{Astrophys. J. Suppl. Ser.}   
\newcommand{\aap}{Astron. Astrophys.}   
\newcommand{\aaps}{Astron. Astrophys. Suppl.}   
\newcommand{\mnras}{Mon. Not. R. Astron. Soc.}   
\newcommand{\nat}{Nature} 

\def\@fnsymbol#1{\ensuremath{\ifcase#1\or *\or \dagger\or \ddagger\or
   \mathsection\or \mathparagraph\or \|\or **\or \dagger\dagger
   \or \ddagger\ddagger \else\@ctrerr\fi}}

\usepackage{setspace}

\begin{document}

\title{\bf Titanium oxide  and chemical inhomogeneity in \\ the atmosphere of the exoplanet WASP-189b} 
\date{}
\author{}
\maketitle
\vspace{-1cm}
\begin{singlespace}
\noindent\author{\textbf{Bibiana Prinoth$^{1\dagger}$,
H.\,Jens Hoeijmakers$^{1}$, Daniel Kitzmann$^{2}$, Elin Sandvik$^{1}$,
Julia\,V. Seidel$^{3,4}$, Monika Lendl$^{3}$, Nicholas\,W. Borsato$^{1}$,
Brian Thorsbro$^{1}$, David\,R. Anderson$^{5}$, David Barrado$^{6}$, Kateryna Kravchenko$^{7}$, Romain Allart$^8$, Vincent Bourrier$^3$,  Heather M. Cegla$^{5,9}$, David Ehrenreich$^{3}$, Chloe Fisher$^{2}$, Christophe Lovis$^{3}$, Andrea Guzm\'{a}n-Mesa$^{2}$, Simon Grimm$^{2}$,
Matthew Hooton$^{10}$, Brett\,M. Morris$^2$, Maria Oreshenko$^{11}$,
Lorenzo Pino$^{12}$, Kevin Heng$^{2,5,13}$}}\\
\vspace{0.01cm}\\
\author{\scriptsize $^1$ Lund Observatory, Department of Astronomy and Theoretical Physics, Lund University, Box 43, 221 00 Lund, Sweden}\\
\author{\scriptsize $^2$ University of Bern, Center for Space and Habitability, Gesellschaftsstrasse 6, 3012 Bern, Switzerland}\\
\author{\scriptsize $^3$ Observatoire astronomique de l’Université de Genève, Chemin Pegasi 51b, 1290 Versoix, Switzerland}\\
\author{\scriptsize $^4$ European Southern Observatory, Alonso de Córdova 3107, Vitacura, Casilla 190001, Santiago, Chile}\\
\author{\scriptsize $^5$ Department of Physics, Astronomy \& Astrophysics Group, University of Warwick, Coventry CV4 7AL, United Kingdom}\\
\author{\scriptsize $^6$ Centro de Astrobiologìa, ESAC Campus, Camino Bajo del Castillo s/n, 28692, Villanueva de la Ca\~nada, Madrid, Spain}\\
\author{\scriptsize $^7$ Max Planck Institute for Extraterrestrial Physics, Giessenbachstraße 1, 85748 Garching, Germany} \\
\author{\scriptsize $^8$ Department of Physics, and Institute for Research on Exoplanets, Université de Montréal, Montréal, H3T 1J4, Canada} \\
\author{\scriptsize $^9$ Centre for Exoplanets and Habitability, University of Warwick, Coventry, CV4 7AL, United Kingdom}\\
\author{\scriptsize $^{10}$ University of Bern, Physics Institute, Division of Space Research \& Planetary Sciences, Gesellschaftsstrasse 6, 3012 Bern, Switzerland}\\
\author{\scriptsize $^{11}$ École Polytechnique Fédérale de Lausanne, PPH 338, Station 13, 1015 Lausanne, Switzerland}\\
\author{\scriptsize $^{12}$ INAF-Osservatorio Astrofisico di Arcetri Largo Enrico Fermi 5, 50125 Firenze, Italy}\\
\author{\scriptsize $^{13}$ Ludwig Maximilian University, University Observatory Munich, Scheinerstrasse 1, Munich 81679, Germany}\\
\author{\scriptsize $^\dagger$ Corresponding author. Email: bibiana.prinoth@astro.lu.se}\\[-1em]
\end{singlespace}
\vspace{1em}
{\bf The temperature of an atmosphere decreases with increasing altitude, unless a shortwave absorber exists that causes a temperature inversion\cite{hubeny2003}. Ozone plays this role in the Earth's atmosphere. In the atmospheres of highly irradiated exoplanets, shortwave absorbers are predicted to be titanium oxide (TiO) and vanadium oxide (VO) \cite{fortney2008unified}. Detections of TiO and VO have been claimed using both low \cite{desert2009tio, sedaghati2017TiO, evans2017statosphere, evans2018optical} and high \cite{nugroho2017high} spectral resolution observations, but later observations have failed to confirm these claims \cite{hoeijmakers2015search,espinoza2019TiO,sedaghati2021TiO} or overturned them \cite{merritt2020non,herman2020search,serindag2021tio}. 
Here we report the unambiguous detection of TiO in the ultra-hot Jupiter WASP-189b \cite{anderson2018wasp} using high-resolution transmission spectroscopy. This detection is based on applying the cross-correlation technique \cite{2010Natur.465.1049S} to many spectral lines of TiO from 460 to 690\,nm. Moreover, we report detections of metals, including neutral and singly ionised iron and titanium, as well as chromium, magnesium, vanadium and manganese (Fe, Fe$^+$, Ti, Ti$^+$, Cr, Mg, V, Mn). 
The line positions of the detected species differ, which we interpret as a consequence of spatial gradients in their chemical abundances, such that they exist in different regions or dynamical regimes. This is direct observational evidence for the three-dimensional thermo-chemical stratification of an exoplanet atmosphere derived from high-resolution ground-based spectroscopy.} \\[-0.5em]

The ultra-hot Jupiter WASP-189b has a high equilibrium temperature of $T_{\texttt{eq}}=2,641 \pm 34$\,K due to its close proximity to its hot A-type host star\cite{anderson2018wasp}. It is one of the brightest transiting planet systems currently known, making it very amenable for spectroscopic studies of its atmosphere. The system is well-characterised, thanks to extensive photometric observations with CHEOPS \cite{lendl2020hot}, including a precise measurement of the orbital parameters, which we adopted in this study. In recent years, similar systems have garnered significant interest within studies that use high-resolution ground-based spectrographs. These have been used to reveal a myriad of absorbing atoms in their emission and transmission spectra \cite{hoeijmakers2018atomic,casasayas2018and,hoeijmakers2019spectral,casasayas2019atmospheric,Cauley2019,ehrenreich2020nightside,gibson2020detection,hoeijmakers2020high,hoeijmakers2020W121,nugroho2020searching,nugroho2020detection,pino2020neutral,stangret2020detection,yan2020temperature,borsa2021atmospheric,tabernero2021espresso}.\\[-0.5em]

We observed time-series of the spectrum of WASP-189 during three transit events, with the High Accuracy Radial velocity Planet Searcher (HARPS) echelle spectrograph at the ESO 3.6 m telescope in La Silla Observatory, Chile (PI: Hoeijmakers, programme number: 0103.C-0472). In addition, we used one archival HARPS transit observation, previously published in Anderson et al. 2018 \cite{anderson2018wasp}, and one archival HARPS-N observation (PI: Casasayas-Barris, programme number: CAT19A\_97). We corrected for telluric absorption using \texttt{molecfit} \cite{smette2015molecfit, kausch2015molecfit}, masked both outliers and spectral regions affected by residuals caused by strong telluric lines, notably O$_2$, and corrected for the RossiterMcLaughlin effect (see Extended Data Figure\,1). We performed cross-correlation analyses \cite{2010Natur.465.1049S} with model spectra of a collection of chemical species, and implemented bootstrap analyses to confirm the statistical robustness of candidate detections (see Methods). We observed transit light-curves using the EulerCam instrument to rule out stellar activity (see Extended Data Figure\,2) and we fit the spectrum to confirm past measurements of metalicity and equatorial rotation velocity (see Extended Data Figure\,3).\\[-0.5em]

Cross-correlation templates were derived from the modelled transmission spectrum of the planet, assuming hydrostatic and chemical equilibrium and an isothermal atmosphere. Each template contains the line opacity of an individual atom, ion or TiO, and the cross-correlation acts to average the spectral lines of each species, weighted by the expected strength of each absorption line as a function of time during the transit event. \\[-0.5em]

Due to the curvature of the exoplanet orbit, absorption signals are Doppler-shifted according to the instantaneous radial velocity of the planet, which scales with the planet's orbital velocity. High-altitude winds from the day- to the night-side act to additionally blue-shift any observed absorption line, while planetary rotation and super-rotational winds result in broadening. This type of analysis can thus be used to measure the orbital velocity of exoplanets, as well as dynamics in their atmospheres \cite{2010Natur.465.1049S, Seager2000,brogi2012signature,Showman2013,Kempton2014,Louden2015, Brogi2016}. We use a template that contains absorption lines of over a hundred atoms and ions to trace the velocity of the planet's atmosphere as it passes through transit, and obtain a best-fit orbital velocity of \vorbprecise\kmpers, consistent with the expected value of  $200.7 \pm 4.9$\kmpers as derived from the orbital parameters determined via precise CHEOPS photometry \cite{lendl2020hot} (see Extended Data Figure\,4).\\[-0.5em] 

For exoplanets that are close to their host stars, the viewing angle varies significantly from the start to the end of the transit event. Co-rotation of the atmosphere with the tidally-locked planet causes absorption lines formed at the leading (morning) or trailing (evening) limbs to be Doppler-shifted in opposite directions. At the same time, there are strong temperature differences between the permanently irradiated day-sides and the cooler night-sides of hot Jovian exoplanets due to tidal locking. Atmospheric chemistry being strongly temperature-dependent, chemical gradients are expected to exist between the two hemispheres\cite{Fortney2006,Spiegel2009,Stevenson2014,Showman2015}. In addition, the atmospheric scale height decreases with decreasing temperature from the day-side to the night-side, especially in the presence of H$_2$ recombination \cite{parmentier2018thermal,pluriel2020}.\\[-0.5em]

\begin{figure*}[h!]
    \begin{center}
    \includegraphics[trim=120 300 100 120, clip,width=0.8\textwidth]{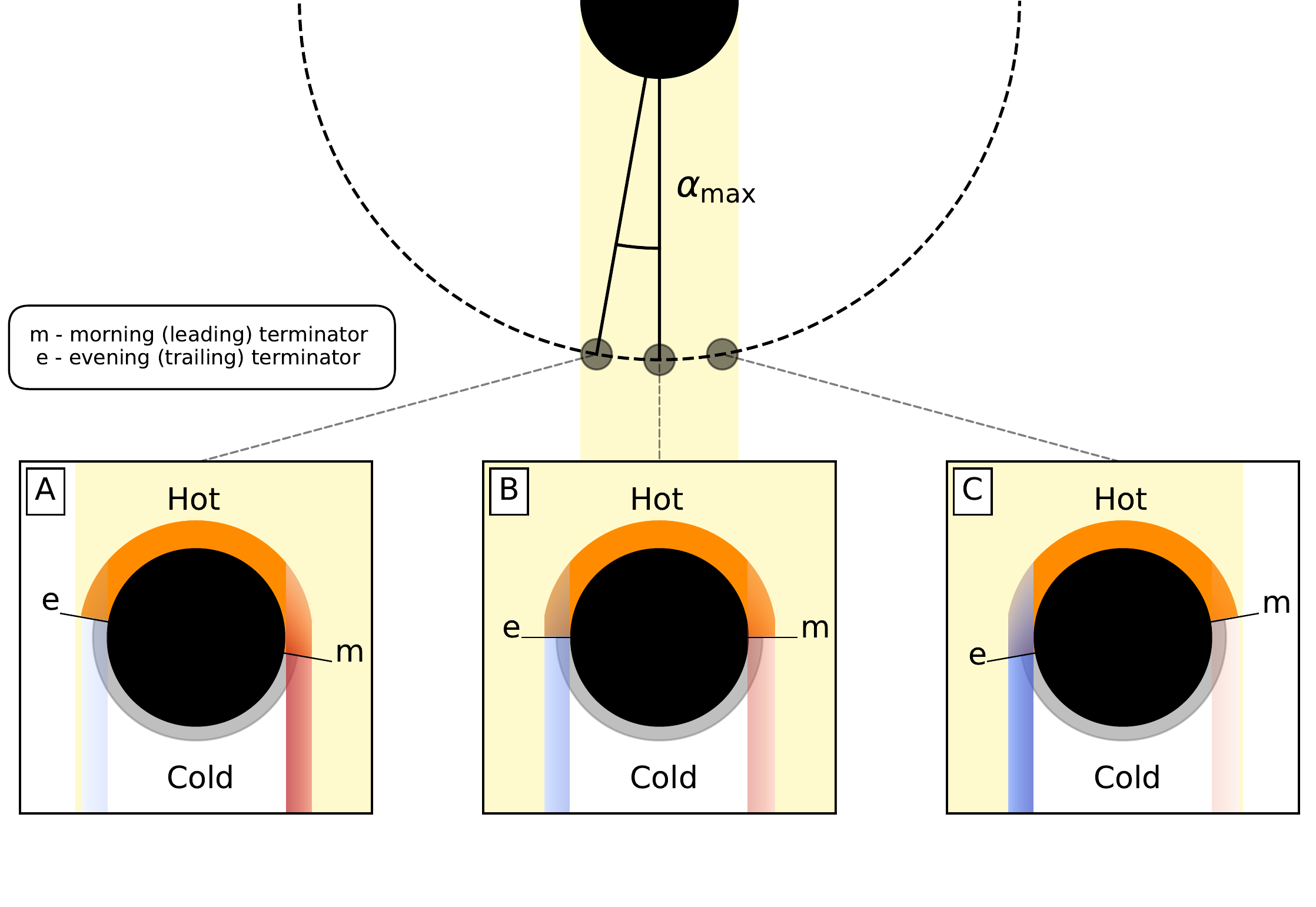}
    \end{center}
    \vspace{-0.2in}
    \caption{\small Schematic showing the contributions of the terminators throughout the course of the transit in the example of a toy planet atmosphere with a hot day-side component in orange, and a cooler night-side component with a significantly reduced scale height in grey (day-night gradient). Due to tidal locking of the planet, the atmosphere is subject to a temperature gradient that alters the atmospheric chemistry as a function of longitude. The schematic shows the contributions to the observed absorption signal originating from the two terminators in the absence of any dynamical effects besides the planetary rotation (no day-to-night-side winds or super-rotational flows). Due to a changing viewing angle ($\alpha_{\rm max} = \arcsin{\left(\sqrt{1-b^2}\frac{R_\ast}{a}\right)}$, where $b$ is the impact parameter, $R_\ast$ is the radius of the star, and $a$ is the semi-major axis), the observed components vary during the transit as follows: \textbf{A:} At the start of the transit, more absorption originates at the morning (leading) terminator because the line of sight passes through more hot atmospheric regions than at the evening (trailing) terminator. At the leading terminator, the absorption spectrum is red-shifted by several kilometres per second because the planet rotates in the counter-clock-wise direction. \textbf{B:} At the centre of the transit, absorption originates from both terminators equally, resulting in a balance between red- and blue-shifted components. \textbf{C:} Towards the end of the transit, absorption at the leading terminator is replaced by absorption at the trailing terminator, which is blue-shifted. The corresponding cross-correlation functions as well as the $K_{\rm p}-V_{\rm sys}$ diagram are shown in the second column of Figure\,\ref{fig:primary} (day-night gradient). \textit{\textbf{Note:}} In absence of a thermal or chemical gradient across the day-to-night-side terminator (i.e. a homogeneous atmosphere, first row of panels in Figure\,\ref{fig:primary}), the two limbs absorb symmetrically and the resulting signal carries neither a net red-  nor a net blue-shift. In the case of a global day-to-night-side wind, the entire absorption signal is blue-shifted (i.e. the third column of panels in Figure\,\ref{fig:primary}).}
    \vspace{-0.1in}
    \label{fig:schematic}
\end{figure*}

Figure\,\ref{fig:schematic} shows a toy-model schematic of how a day-to-night-side temperature gradient is expected to alter the observed cross-correlation signatures\cite{ehrenreich2020nightside} -- in the absence of any atmospheric dynamics apart from co-rotation with the tidally-locked planet. At the beginning of the transit event, more absorption originates at the leading (morning) terminator, because the line of sight passes through hotter atmospheric regions than at the trailing (evening) terminator. At the leading terminator, the atmosphere is red-shifted because the planet rotates in the counter-clock-wise direction, causing an effective redshift of the observed absorption signal. Towards the end of the transit, absorption at the leading terminator is replaced by absorption at the trailing terminator, which is blue-shifted. This effect manifests itself as a decrease in the inferred orbital velocity of the planet, which is a primary observable in analyses of high-resolution spectroscopy of exoplanets. Without a thermal or chemical gradient across the day-to-night-side terminator, the two limbs absorb symmetrically and the observed orbital velocity is equal to the true orbital velocity of the planet. \\[-0.5em]

\begin{figure*}[h!]
    \begin{center}
    \includegraphics[trim=20 220 20 160, clip, width=\textwidth]{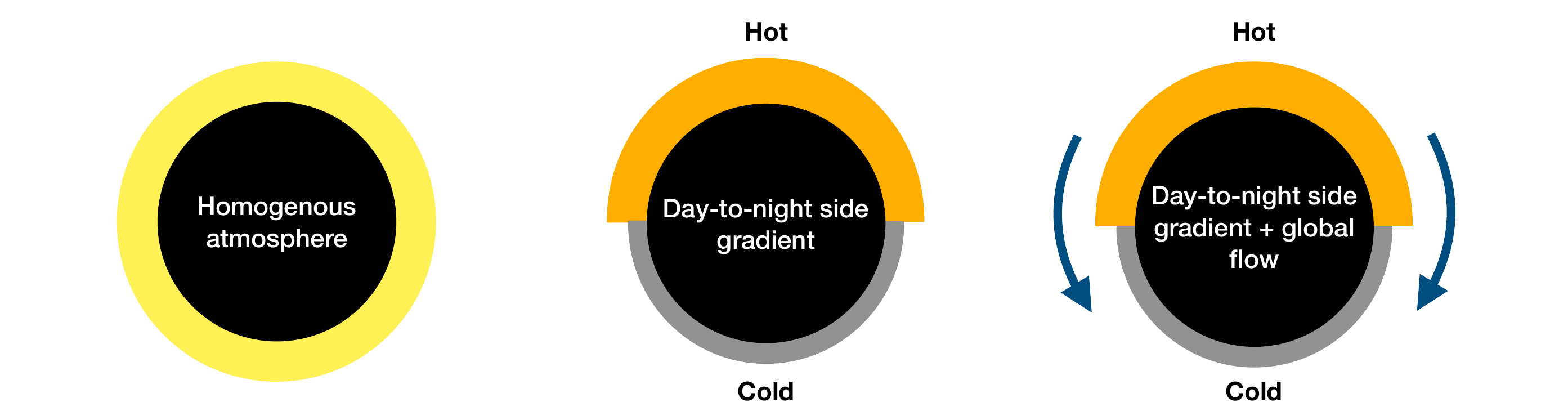}
    \includegraphics[trim=60 43 30 75, clip, width=\textwidth]{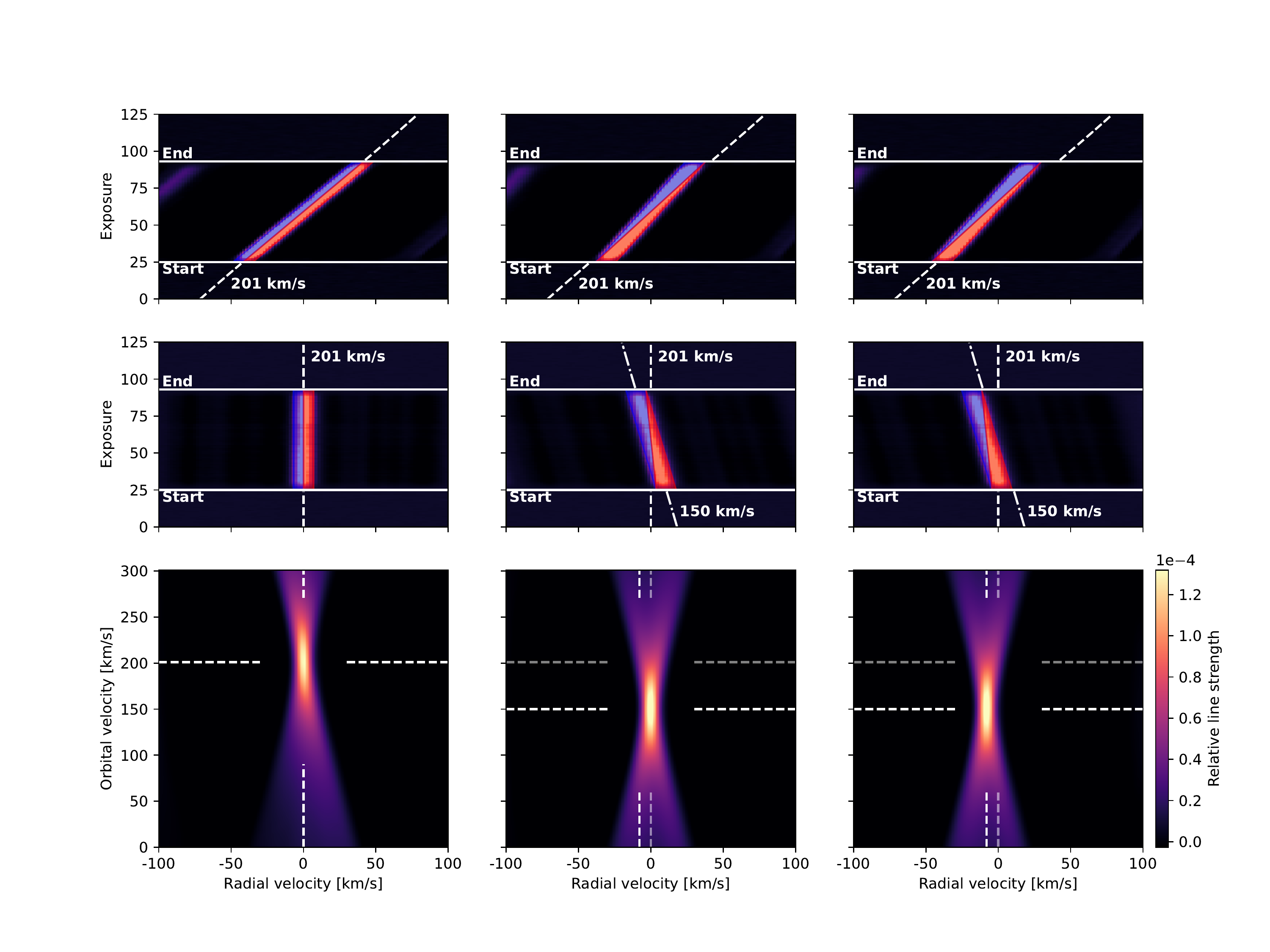}
    \end{center}
    \vspace{-0.2in}
    \caption{\small Expected cross-correlation functions and velocity-velocity diagrams for three toy-models of the planet atmosphere. \textbf{Row 1, left:} Expected behaviour for three toy-models of the planet atmosphere for a homogeneous atmosphere without any flows, temperature or chemical gradients. \textbf{Middle:} A general day-to night side gradient results in a decreasing scale-height towards the night-side, indicated by the orange (hot) and grey (cold) parts of the atmosphere with changing contributions from the blue- (trailing limb) and red-shifted (leading limb) components. For this toy-planet it is assumed that the gradient in the tested species causes an apparent an orbital velocity of 150 \kmpers. \textbf{Right:} A general day-to-night side gradient and global flow, introducing a wind from the hot day side to the cold night side. The global wind causes a shift in radial velocity of 8 \kmpers (right)
    \textbf{Row 2:} Time-dependent cross-correlation signal over the course of the transit in the stellar rest-frame. The contribution of the limbs are indicated in blue (evening) and red (morning). 
    \textbf{Row 3:} Time-dependent cross-correlation signal shifted to the orbital velocity of 201\kmpers. For a homogeneous atmosphere, the signal appears as a symmetrical, vertical feature. In the case of global day-to-night side gradients, the signal is tilted, corresponding to a lower orbital velocity. Adding global winds, shifts the the whole signature to lower radial velocities.
    \textbf{Row 4:} $K_{\rm p}-V_{\rm sys}$ diagram showing the expected and observed orbital and radial velocities. In case of different velocities, the observed velocity is indicated with an opaque line. In the case of global day-to-night side gradients, absorption from the red-shifted leading limb dominates at the start of the transit, and vice versa. This leads to an apparent misalignment when the signal is co-added in the planet's rest-frame, and the signal therefore appears to maximise at an orbital velocity that is lower than the true orbital velocity. A global day-to-night-side flow acts to blue-shift the entire signal towards negative radial velocities in addition to asymmetries related to day-to-night-side gradients in temperature or chemistry.}
    \label{fig:primary}
\end{figure*}

Atmospheric dynamics other than atmospheric co-rotation alter this schematic in two ways. Firstly, flows act to shift or deform the observed absorption lines due to the Doppler effect. Super-rotational winds cause an exacerbation of the red- and blue-shifts of the two limbs, and an apparent broadening of the observed spectral lines. An over-all day-to-night-side flow, as expected to be an important dynamical regime for highly irradiated gas giants \cite{Showman2013}, causes a blue-shift of the entire signature towards negative radial velocities. Secondly, winds transport heat around the globe, changing the global temperature distribution. For hot Jupiters, this commonly leads to an offset between the hottest point on the day-side of the planet and the sub-stellar point. This affects the transmission spectrum, as was recently observed in the hot Jupiter WASP-76b\cite{ehrenreich2020nightside}. The abundance of Fe on the day-side was inferred to be centred on a point west-ward of the sub-stellar point, based on stronger observed blue-shifted absorption at the trailing limb, and a total disappearance of red-shifted absorption from the leading limb by the time of transit centre, implying that Fe condenses out of the gas phase on the planet night-side. The effects on the signature in the cross-correlation functions and in the $K_{\rm p}-V_{\rm sys}$ diagrams were recently explored by Wardenier et al, 2021\cite{Wardenier2021} and are explained in Figure\,\ref{fig:primary}.\\

Our analysis of the HARPS and HARPS-N observations of the WASP-189 system resulted in strong detections of 9 species (Fe, Cr, Mg, Mn, Ti, V, Fe$^+$, Ti$^+$ and TiO), as well as tentative detections of 5 species (Na, Ca, Sc$^+$, Cr$^+$ and Ni) (see Table \ref{tab:best_fit_gauss}, Figure\,\ref{fig:secondary}, Extended Data Figures\,5 and 6, and Methods). For most neutral atoms, we observed line depths that are consistent with models that assume local thermodynamic equilibrium (LTE) and hydrostatic and chemical equilibrium, as revealed by model injection (see Extended Data Figure\,7). Strong absorption by metal ions is inconsistent with this class of models, as these are not predicted to be observable. Such departures from model predictions suggest that non-LTE effects, hydrodynamic escape or night-side condensation may be important for ultra-hot Jupiters\cite{Fossati2020,lothringer2020uv}. \\

Some species show a significant blue-shift compared to the systemic velocity of $-24.452 \pm 0.012$\kmpers \cite{anderson2018wasp}, indicative of day-to-night-side flows, as is commonly observed in hot gas-giant exoplanets \cite{2010Natur.465.1049S,hoeijmakers2018atomic,hoeijmakers2020high,hoeijmakers2020W121,stangret2020detection,Brogi2016, casasayas2019carmenes,bourrier2020hot}. These shifts are not all consistent with each other, suggestive of differences in the prevailing wind patterns at the locations in the atmosphere at which the various species absorb. Besides blue-shifts, we observe that to place the absorption lines of certain species at their rest-frame velocities, these need to be Doppler-shifted assuming values of the orbital velocity $K_p$ that are smaller than the true orbital velocity of \vorb\kmpers, in particular Mg and Cr. As recently described by Wardenier et al. 2021\cite{Wardenier2021}, deviations between the apparent orbital velocity and the true planet rest-frame are the result of a day-to-night-side gradient in chemical abundance driven by the temperature contrast between the two hemispheres leading to night-side condensation, combined with asymmetries in the wind speeds between the morning and evening limbs \cite{ehrenreich2020nightside}. This interpretation, however, depends on the orbital velocity of the planet as derived using the stellar radius $R_*$, which was modelled from the stellar spectrum \citep{lendl2020hot}. In the event that this value of the stellar radius is systematically overestimated by 10\%, the apparent values of $K_p$ at which the cross-correlation signatures are maximising could coincide with the actual orbital velocity of the planet. Nevertheless, from the observation that different species absorb at different systemic velocities, we conclude that the global distributions of the observed species are not all equal, but instead vary significantly between atoms, ions and molecules (TiO).\\

Our observations of WASP-189b imply that the absorption of detected species originates in different thermal, chemical and/or dynamical regimes, and that asymmetric absorption is not purely due to a rapidly decreasing scale height near the terminator region, which would affect all species uniformly. Offsets in orbital velocity caused by day-night abundance gradients and wind-speed asymmetry \cite{ehrenreich2020nightside,Wardenier2021}, if real, appear relatively strongest for Mg and Cr. However, absorption signals of Ti and TiO do not show similar offsets even though the condensation temperature of Ti is significantly higher \cite{lodders2003abun}. This could mean that Mg and Cr condensation is limited to latitudinal regions or altitudes where Ti and TiO are not present (i.e. pressures below $\sim\,10^{-5}$ bar, see Extended Data Figure\,8), or that Ti/TiO and Mg/Cr both condense out, but that titanium is remixed more quickly near the morning terminator -- implying that these species predominantly exist in different dynamical regimes. Similarly, radial velocity offsets (see $v_0$ in Table\,1) differ significantly between some species, again implying differences in the atmospheric dynamics prevalent at the locations at which these species absorb. These observations therefore empirically demonstrate that the atmospheres of intensely irradiated gas-giant exoplanets must indeed have non-trivial three-dimensional structures, and that these significantly impact observables in high-resolution transit transmission spectra, the details of which are to be elaborated using Global Circulation Models (GCM) simulations \cite{Wardenier2021}. As put forth previously \cite{pluriel2020, Mendonca2018}, it is expected from GCMs that temperature, chemical abundances and wind speeds indeed vary strongly as a function of latitude around the terminator region. In the case of WASP-189b, dynamical heterogeneity may be intensified by variability in the effective irradiation, due to its polar orbit around its oblate and gravity-darkened host star \cite{anderson2018wasp, lendl2020hot}, as has recently been observed in KELT-9\,b\cite{Ahlers2020}.\\[-0.5em]

Furthermore, these observations constitute a detection of the TiO molecule in the transmission spectrum of WASP-189b with a 5.6$\sigma$ detection confidence (see Table \ref{tab:best_fit_gauss}). Previously claimed detections of the TiO molecule in exoplanet atmospheres have predominantly been made at low spectral resolution\cite{desert2009tio,sedaghati2017TiO,evans2018optical,Changeat2021,Chen2021} -- where it is difficult to discern unambiguously from other broad-band absorbers or systematic noise. The only detection that was made using high resolution spectra relied on observations of the day-side of WASP-33b during a single epoch\cite{nugroho2017high}. A later re-analysis of the same data \cite{serindag2021tio} and further observations of the same object \cite{herman2020search} have failed to confidently reproduce that signal. \\[-0.5em]

In light of the past ambiguity regarding the TiO detection in WASP-33b, we have analysed subsets of our data, alternately excluding one of the five transit time-series. For each subset, TiO is recovered at an average detection significance of 4.9$\sigma$, confirming that the signal originates from all five transit observations combined. Additionally, we recovered the signal for each time-series individually, with an average detection significance of 3.8$\sigma$ for each time-series covering the full transit. This is the first unambiguous detection of TiO in the transmission spectrum of an exoplanet, and may be the first robust detection of this molecule in the atmosphere of any exoplanet entirely. Importantly, we conclude that even for planets at high temperatures such as WASP-189b, TiO may be a significant source of stratospheric heating, in addition to atomic metals that absorb efficiently at short wavelengths \cite{lothringer2020uv}.\\[-0.5em]

As more ultra-hot Jupiters are being observed with sensitive high-resolution spectrographs on large ground-based telescopes and space-based observatories such as the James Webb Space Telescope, our observations empirically demonstrate that the successful interpretation of observations of this type of planet requires that the theory of exoplanet atmospheres appreciates the three-dimensional nature of these atmospheres and that insights derived from GCMs, atmospheric chemistry and radiative transfer are unified. Observations of hot Jupiters have exhausted the flexibility of one-dimensional models, providing strong motivation for innovations in data analysis techniques, numerical modelling and fundamental atmospheric theory.\\[-0.5em]

\begin{table*}[ht]
    \begin{center}
    \begin{tabular}{l|lllll}
    \hline
    \hline
         & A [$\times 10^{-6}$] & $v_0\ [{\rm km / s}]$ & FWHM $[{\rm km / s}]$ & $v_{\rm orb, ext}/K_{\rm p} [{\rm km / s}]$ &  $\sigma$\\
    \hline
    Cr & $45.8 \pm 5.7$ & $-29.03 \pm 0.75$ & $12.1 \pm 1.8$ &  $ 159.2 \pm 7.8 $ & $8.0$ \\ 
    Fe & $75.0\pm 4.3$ & $-28.09 \pm 0.43$ & $15.3 \pm 1.0$ & $ 191.8 \pm 3.9 $ & $17$ \\ 
    Fe$^+$ & $228\pm 17$ & $-25.56 \pm 0.51$ & $14.1 \pm 1.2$ & $ 189.4 \pm 6.4 $ & $14$ \\ 
    Mg & $177\pm 22$ & $-24.3 \pm 1.2$ & $20.5 \pm 2.9$ & $ 167 \pm 10 $ & $8.1$ \\ 
    Mn & $83\pm 13$ & $-30.6 \pm 1.3$ & $17.2 \pm 3.1$  & $202 \pm 12$ & $6.5$ \\  
    Ti & $38.4\pm 4.4$ & $-24.90 \pm 0.85$ & $15.1 \pm 2.0$ & $ 202.1 \pm 9.0 $ & $8.7$ \\ 
    Ti$^+$ & $139\pm 14$ & $-23.95 \pm 0.81$ & $15.9 \pm 1.9$ & $ 180.9 \pm 8.1  $ & $9.7$ \\ 
    TiO & $1.67\pm 0.30$ & $-28.53 \pm 0.95$ & $10.9 \pm 2.3$ & $ 185.68 \pm 10.16$ & $5.6$ \\
    V & $30.6 \pm 4.9$ & $-28.3 \pm 1.7$ & $21.7 \pm 4.0$ & $218 \pm 13$ (201) & $6.2$ \\ 
    \hline
    Ca$^\ast$ & $38.4\pm 8.6$ & $-28.6 \pm 1.9$ & $16.6 \pm 4.4$ & \vorb & $4.4$ \\ 
    Cr$^+$ & $121\pm 31$ & $-29.0 \pm 1.1$ & $8.5 \pm 2.5$ & \vorb & $3.9$ \\ 
    Na$^\dagger$ & $49\pm 15$ & $-29.4 \pm 1.4$ & $9.0 \pm 3.2$ & \vorb & $3.2$ \\ 
    Ni & $50\pm 11$ & $-24.0 \pm 1.6$ & $14.1 \pm 3.7$ & \vorb & $4.4$ \\ 
    Sc$^+$ & $122\pm 41$ & $-20.2 \pm 1.0$ & $7.1 \pm 3.0$ & \vorb & $3.0$ \\ 
    \hline
    \end{tabular}
    \caption{Best-fit parameters of Gaussian fits to the one-dimensional cross-correlation functions obtained by combining all five nights of observation. $A$ corresponds to the best-fit line depth of the absorbing species above the spectral continuum. $v_0$ corresponds to the radial velocity of the line centre, as measured in the rest-frame of the solar system. The real systemic velocity is $-24.452 \pm 0.012$\kmpers \cite{anderson2018wasp}, meaning that most species are observed to have a significant blue-shift. FWHM denotes the Gaussian Full-Width at Half-Maximum, and $\sigma$ indicates the confidence of the detection assuming Gaussian standard deviations. Ca, Cr$^+$, Na, Ni and Sc$^+$ are classified as tentative, due to relatively low detection significance ($<5.0\sigma$). $v_{\rm orb, ext}$ denotes the orbital velocity at which the signal maximises according to the $K_{\rm p}-V_{\rm sys}$ diagram as obtained by model fitting, see Methods. Except for V, all detected signals are extracted at the best fit orbital velocity denoted $v_{\rm orb, ext}$. V is extracted at \vorb\kmpers.  $^\ast$The signal of Ca is irregularly extended, with a relatively large uncertainty on the best-fit orbital velocity, see Extended Data Figure\,2. As Ca is classified as a tentative detection, we allow the 1D cross-correlation function to be extracted at the true orbital velocity of the planet instead. $^\dagger$The signal of Na is obtained using the cross-correlation technique. We also performed a targeted Na analysis resulting in a higher detection significance, see Methods.}
    \label{tab:best_fit_gauss}
    \end{center}
\end{table*}

\newpage

\newpage
\newpage

\begin{figure*}[h!]
    \begin{center}
    \includegraphics[width=\textwidth]{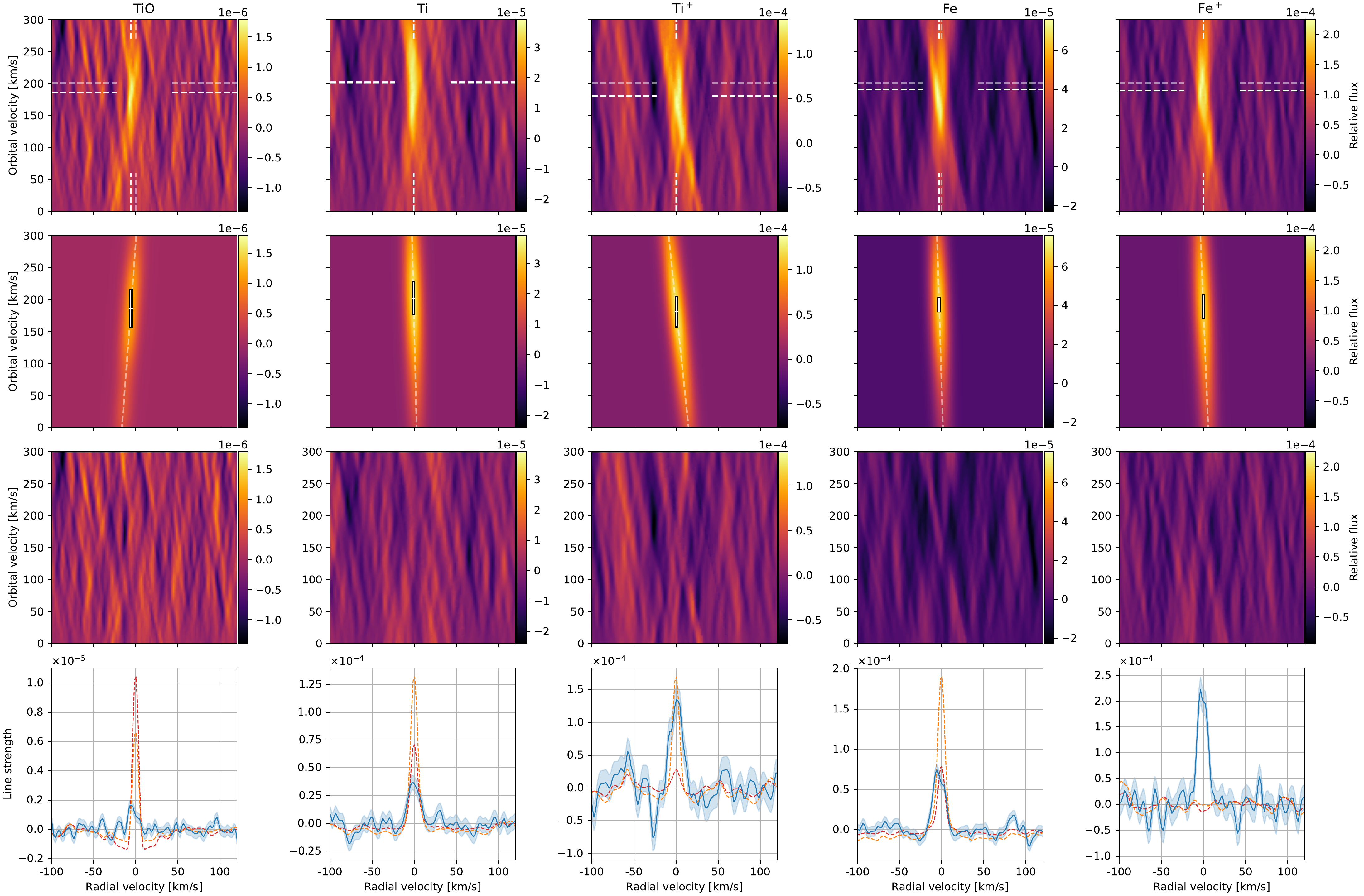}\\
    \vspace{1cm}
    \includegraphics[width=0.8\textwidth]{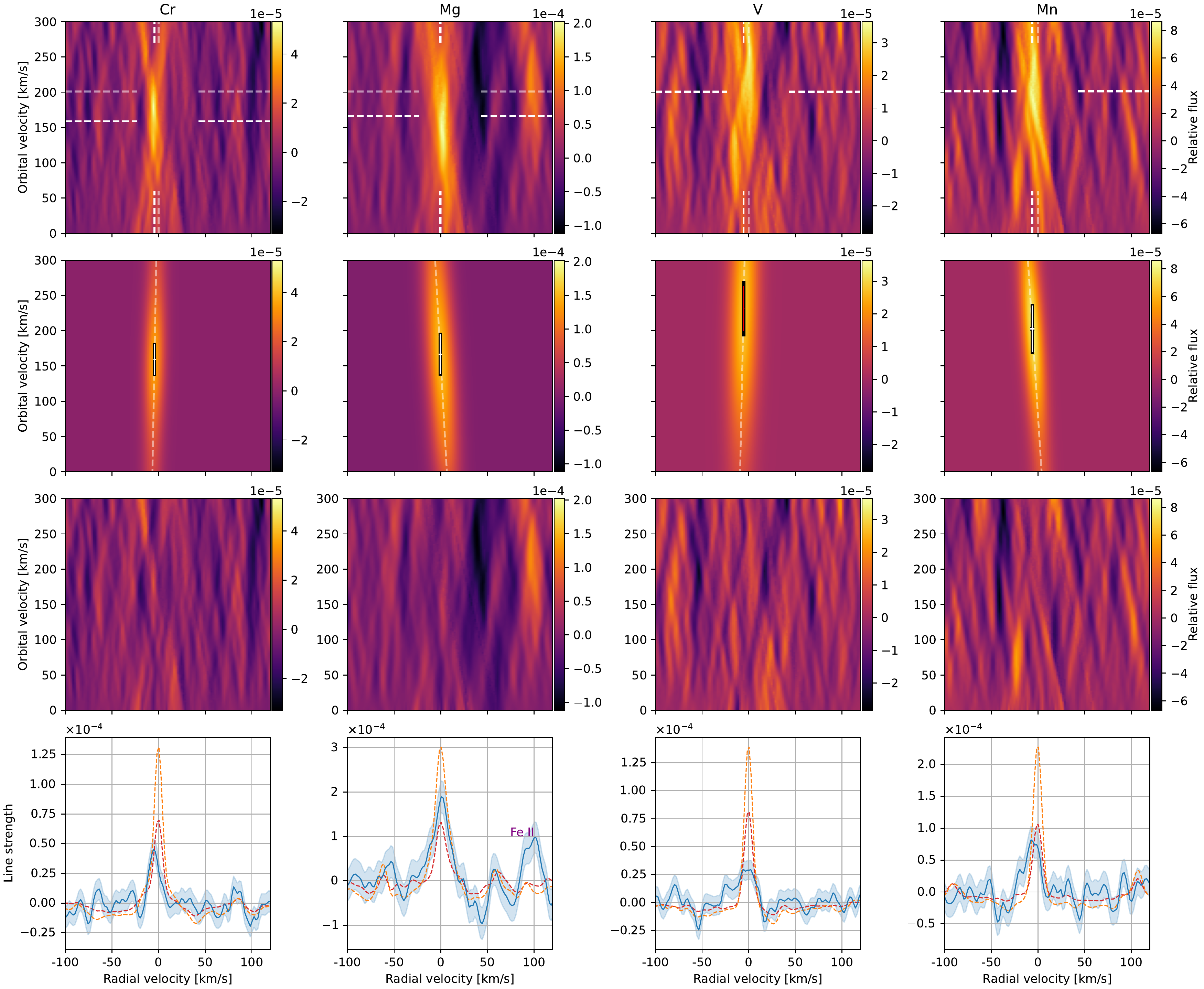}
    \end{center}
    \vspace{1cm}
\end{figure*}
    \captionof{figure}{\small Overview of the detections of TiO, Ti, Ti$^+$, Fe, Fe$^+$, Cr, Mg, V and Mn. \textbf{Rows 1\&5:} Velocity-velocity ($K_{\rm p}-V_{\rm sys}$) diagrams showing the absorption signals of the detected species in the rest frame of the star. The horizontal opaque dashed lines indicate the orbital velocity at which the signals were extracted, see also Table\,\ref{tab:best_fit_gauss}. \textbf{Rows 2\&6:} Best-fit two-dimensional Gaussian models to the signatures visible $K_{\rm p}-V_{\rm sys}$ diagrams in top panels, see Methods. The best-fit orbital velocity including 3$\sigma$ uncertainties are shown in the panels as white bars. The uncertainty for V corresponds to 1$\sigma$. Dashed lines indicate the slopes of the best fit. \textbf{Rows 3\&7:} Residual $K_{\rm p}-V_{\rm sys}$ diagrams after subtraction of the best-fit two-dimensional Gaussian models. 
    \textbf{Rows 4\&8} Cross-correlation functions stacked in the rest-frame according to the best fit orbital velocities stated in Table \ref{tab:best_fit_gauss}. The shaded region indicates the expected $1 \sigma$ uncertainty. Dashed lines show expected signal strengths, obtained by injecting and recovering the signatures of model spectra, assuming isothermal atmospheres at 2,000\,K (red) and 3,000\,K  (orange) respectively.
    The strong departure from model predictions of the cross-correlation function of Fe$^+$ suggests that non-LTE effects, hydrodynamic escape or night-side condensation may be important for ultra-hot Jupiters\cite{Fossati2020,lothringer2020uv}.\label{fig:secondary}}

\newpage
\section*{Methods} 
\subsection*{Observations, data reduction and telluric correction}
Transit observations with HARPS were performed in the nights of April 14, April 25 and May 14, 2019. Earlier observations covering a partial transit were performed in the night of March 26, 2018 and have previously been published\cite{anderson2018wasp}. A fifth night of observations covering a full transit was performed in the night of May 6, 2019, using HARPS-N. A log of the observations is provided in Supplementary Data Table\,1.\\

All science observations are performed with fibre A on the target and fibre B on the sky. Due to weather circumstances, observations on the night of 2019-04-25 started when the planet was already in transit and hence offer no baseline before the transit event. Similarly, the night of Anderson et al.\,(2018)\cite{anderson2018wasp}, covers only half a transit.\\

All HARPS observations were reduced by the HARPS Data Reduction Software (DRS) v3.8. Earth's atmospheric telluric lines were removed using \texttt{molecfit} \cite{smette2015molecfit, kausch2015molecfit}. \texttt{Molecfit} was applied to the one-dimensional spectra generated by the reduction pipeline to build a model for the telluric transmission spectrum of the entire wavelength range covered by HARPS. This model was run for each exposure in each time-series and for all nights individually. Because the HARPS spectra are given in the barycentric rest-frame of the Solar system, we used the Earth barycentric radial velocity (BERV) values to shift each exposure back into the rest-frame of the instrument. This step is necessary to ensure correct modelling of the telluric lines with \texttt{molecfit}. Regions containing strong H$_2$O and O$_2$ absorption lines around 595, 630 and 647.5\,nm were used to fit the model, selecting small wavelength ranges that contain telluric lines surrounded by a flat continuum where no stellar lines are present. The thus obtained telluric models were then interpolated onto the same wavelength grid as the individual spectral orders of HARPS and finally divided out to remove telluric effects.\\

In addition to spectral observations with HARPS, we obtained simultaneous photometric observations with the EulerCam instrument on the Euler-Swiss telescope at La Silla Observatory, during two of the four transit events considered in this study, on the nights of 14-04-2019 and 24-04-2019. Although not expected for hot A-stars, these observations were aimed at detecting photometric variability, e.g. due to stellar spot crossings or outbursts. The data were reduced using standard methods \cite{lendl2012}. Extended Data Figure\,2 shows the phase-folded light-curve, fit with a standard model \cite{MandelAgol}, multiplied by a baseline model consisting of a 2$^{\textrm{nd}}$ order polynomial in the target peak count level and a linear trend in time, using CONAN\cite{lendl2020CONAN}. No variability of astrophysical origin was identified, and later observations by CHEOPS have also not revealed significant variability in this host star \cite{lendl2020hot}. 

\subsection*{Preparatory corrections}
After removal of telluric lines, further preparatory corrections were made. Individual spectra were Doppler-shifted to the rest-frame of the host star by correcting for the Earth's velocity around the barycentre of the Solar system as well as the radial velocity of the star caused by the orbiting planet, so that the stellar spectrum had a constant velocity shift consistent with the systemic velocity of $\sim \,-26$\kmpers. Following Hoeijmakers et al.\,(2020) \cite{hoeijmakers2020W121}, we rejected 5$\sigma$-outliers from the spectral time-series by applying an order-by-order sigma-clipping algorithm with a running median absolute deviation computed over sub-bands of the time-series with a width of 40 pixels. We further manually flagged spectral columns with visible systematic noise caused mainly by residuals of deep telluric lines. For any of the nights, this did not affect more than 1.6\% of the total number of spectral pixels in each time-series. In the case of TiO, only selected wavelength regions beyond 460\,nm were included where the line list is known to be relatively accurate \cite{mckemmish2019exomol}.  The excluded wavelength ranges were up to 460 nm, 507.2-521.6 nm, 568.8-580.6 nm, 590.9-615.4 nm and 621.0 - 628.0 nm. Also following Hoeijmakers et al.\,(2020) \cite{hoeijmakers2020W121}, we performed a colour correction by fitting low-order polynomials to the residuals of each order after dividing by their time-average, thereby equalising the shape of the broad-band continuum of each exposure, order-by-order. This removed any colour-dependent variations in the illumination (including variations in the blaze function), and hence of the effective response of the spectrograph.

\subsection*{Cross-correlation templates}
Cross-correlation templates were constructed using model transmission spectra of individual atoms as well as TiO assuming an isothermal atmosphere of 2,000\,K for TiO, 3,000\,K for neutral atoms, or 4,000\,K for ions, because most of these are not expected to absorb significantly at temperatures below 4,000\,K. In addition, we constructed a template containing 131 sources of line opacity at a temperature of 3,000\,K for the purpose of measuring the instantaneous radial velocity of the planet's atmosphere. The templates were broadened to approximately match the line-spread function of the HARPS instrument, with a full-width at half maximum of 2.7\kmpers, to avoid under-sampling. 

The spectra were cross-correlated with the ensemble of templates, producing cross-correlation coefficients that are effectively weighted averages of the spectral lines included in the templates. The cross-correlation coefficients are given by

\begin{equation}
    C(v,t) = \sum_{i = 0}^{N} F_{\rm i}(\lambda, t)T_i(v)
    \label{eq:cc}
\end{equation}

where $F_{\rm i}(\lambda, t)$ are the spectra of the time-series, i.e. all spectral points in all echelle orders of the spectrum obtained at a given time $t$, $T_i(v)$ are the corresponding values of the template Doppler-shifted to a radial velocity $v$. $T(v)$ takes non-zero values within the spectral lines of interest and is normalised such that \(\sum_{i=0}^N T_i(v) = 1\). This procedure generates the two-dimensional cross-correlation functions for each night of observations and each species. The uncertainty intervals are determined through Gaussian error propagation of the expected photon noise on the individual spectra.

\subsection*{Cleaning steps}

\subsubsection*{Removal of the Doppler shadow}\label{sec:doppler}
During transit the planet partially obscures areas of the rotating star. This obscured area introduces residual spectral lines when performing differential transmission spectroscopy, which create a spurious cross-correlation signal often called the Doppler shadow. Generally this feature occurs at an apparent radial velocity that is different from the planetary atmosphere, and can thus be removed without significantly affecting the planetary signature. This is especially true for WASP-189b, which resides on a polar orbit \cite{anderson2018wasp}, resulting in a Doppler shadow residual that is nearly constant in radial velocity, as also observed for KELT-9b \cite{hoeijmakers2018atomic, gaudi2017giant}. We constructed empirical models of the Doppler shadow for neutrals as well as ions using the cross-correlation functions of Fe and Fe$^+$ respectively by fitting a Gaussian profile of which the centroid velocity is prescribed following Cegla et al. 2016\cite{cegla2016rossiter}, while the amplitude {and} width are allowed to vary according to low-order polynomials, to capture variations related to limb darkening and gravity darkening of the host star. We fitted a second Gaussian component to correct for the wide, negative pseudo-absorption inherent to the Rossiter-McLaughlin effect in normalised spectra. These two components form a model that is subtracted from the cross-correlation function of each of the species, multiplied by a scaling factor in a least-squares manner, see Extended Data Figure\,1 and Supplementary Data Figure\,1, following Hoeijmakers et al. (2020)\cite{hoeijmakers2020W121}. During this removal we protected the planetary signature by masking out the radial velocity range of the planet at each orbital phase when fitting the Doppler shadow. 

\subsubsection*{Detrending correlated noise and aliases}
\label{sec:detrending}
After removing the Doppler shadow, some correlated structure was still present in the two-dimensional cross-correlation functions between the times of ingress and egress, mainly caused by aliases between strong lines in the cross-correlation templates and the stellar absorption lines obscured by the transiting planet. Because the planet is on a near-polar orbit, these aliases take the form of near-vertical structures in the cross-correlation functions. To remove these, and any other systematic noise constant in radial velocity as a function of time, we fitted and subtracted a polynomial of degree one at each column of the parts of the two-dimensional cross-correlation functions that correspond to in-transit exposures. In addition, we perform a Gaussian high-pass filter with a width of 100\kmpers to remove broad-band structures in the spectral direction\cite{hoeijmakers2018atomic,hoeijmakers2019spectral,hoeijmakers2020W121}. The results of these cleaning steps are shown in Extended Data Figure\,1 and Supplementary Data Figure\,1.

\subsection*{Shift into the rest-frame of the planet and fitting}
\label{shift_restframe}
The resulting detrended, cleaned, two-dimensional cross-correlation functions were shifted towards the expected rest-frame of the planet, assuming a value for the orbital velocity. To make sure that no Doppler shadow removal residual unintentionally adds to the signal of the planet, we masked out the overlapping region at the end of the transit. To co-add in-transit cross-correlation functions, they were weighted according to the mean flux of their corresponding spectra, yielding a flux-weighted, time-averaged one-dimensional cross-correlation function for each assumed orbital velocity. This resulted in maps of the cross-correlation of each species in the orbital velocity / systemic velocity space ($K_{\rm p}-V_{\rm sys}$) where the peak of the planetary signal is expected to be located at the true orbital and systemic velocities in the absence of atmospheric dynamics. For combining the cross-correlation functions of the four independent transits we averaged the $K_{\rm p}-V_{\rm sys}$ diagrams by similarly weighing them according to the total in-transit flux recorded by HARPS(N) during each observation. \\

We performed an analysis to extract the line shape, depth and location in $K_{\rm p}-V_{\rm sys}$ space, by fitting a 2-dimensional Gaussian model that allows for correlation between the orbital and systemic velocity via a rotation parameter, to the signature in $K_{\rm p}-V_{\rm sys}$ space (see Figure \ref{fig:secondary}), with parameters described by low-order polynomials to approximate the flaring shape of the signatures that is characteristic of transit time-series. This model is evaluated at steps of 25\kmpers in $K_{\rm p}$, to diminish the strong correlation between values in the vertical direction of the $K_{\rm p}-V_{\rm sys}$ diagram: In our time-series, a change in $K_{\rm p}$ of 25\kmpers causes a relative shift of 2\kmpers in exposures taken 45 minutes apart. Given that the cross-correlation function is evaluated in steps of 2\kmpers to eliminate correlation between adjacent cross-correlation function velocity steps; and that the transit transit duration is 4.3 hours; two rows in $K_{\rm p}-V_{\rm sys}$ space that are separated by 25\kmpers are constructed from cross-correlation function samples of which more than 80\% are unique. Because the templates were broadened to a full width at half maximum of 2.7\kmpers and the cross-correlation analysis was performed in steps of 2\kmpers, only every second data point (i.e. every 4\kmpers) and its corresponding uncertainty were used to fit the Gaussian profile in order to mitigate correlations between neighbouring cross-correlation points. 
We therefore treat 25\kmpers steps in $K_{\rm p}$ and 4\kmpers steps in radial velocity as statistically independent from each other, allowing us to fit a two dimensional model to the signatures in the $K_{\rm p}-V_{\rm sys}$ diagram while assuming independently normally distributed uncertainties.
At these observed orbital velocities of the signals (see Table \ref{tab:best_fit_gauss}), we extracted the one-dimensional cross-correlation function and fitted a Gaussian to the peak of the cross-correlation function at the location of the systemic velocity to measure the line depth of the detected absorption. The fitting results are shown in Table \ref{tab:best_fit_gauss}. Even though candidate signals of Ca, Cr$^+$, Na, Ni and Sc$^+$ are detected with confidence greater than 3$\sigma$, we conservatively choose to classify these as tentative, based on them having anomalous centre positions or widths, or extended shapes in $K_{\rm p}-V_{\rm sys}$ space. All species classified as tentative have formal confidence levels less than 5$\sigma$, while all claimed detections are at levels of $5\sigma$ or greater. Compared to the systemic velocity of $-24.452 \pm 0.012$\kmpers\cite{anderson2018wasp} most species are significantly blue-shifted, even if the radial velocity of Anderson et al.\,(2018)\cite{anderson2018wasp} suffered from larger than expected uncertainties, at the level of approximately 0.1\kmpers. This indicates the presence of a day-to-night-side wind on the level of several \kmpers, similar as been observed in other hot Jupiters \cite{2010Natur.465.1049S, casasayas2019atmospheric, ehrenreich2020nightside, Brogi2016, bourrier2020hot}.

\subsection*{Model injection}
\label{modelinjection}
Following the procedure in Hoeijmakers et al.\,(2019)\cite{hoeijmakers2019spectral}, modelled transmission spectra for WASP-189b were injected at two temperatures (2,500 K, close to the planetary equilibrium temperature of 2,641 K \cite{anderson2018wasp}, and 3,000 K, between the equilibrium temperature and the day-side temperature of 3,400\,K \cite{lendl2020hot}), assuming the planet's atmosphere to be isothermal, in chemical and hydrostatic equilibrium and of solar metallicity. The planetary parameters such as the planetary radius and the surface gravity were adopted from Lendl et al.\,(2020) \cite{lendl2020hot} (see also Extended Data Figure\,4), assuming a reference pressure of 10 bar at a radius of 1.619\,$R_{\rm Jup}$. We used \texttt{FastChem} \cite{stock2018fastchem} to calculate the chemical abundance profiles and followed the radiative transfer procedure as performed in Gaidos et al.\,(2017) \cite{gaidos2017exoplanet}. Opacity functions of 128 atomic neutrals and individual ions were included in this model, as well as H$_2$O, TiO and CO. These were computed using the open-source \texttt{HELIOS-K} opacity calculator \cite{grimm2015helios,Grimm2021} from line lists provided by VALD and Exomol for atoms and molecules, respectively \cite{mckemmish2019exomol, Li2015,ryabchikova2015major, tennyson2016exomol, Polyansky2018}.

The model spectra are full forward models, including continuum and accurate line depths and profiles. No additional normalisation is performed for the purpose of injection. The modelled transmission spectra were injected into the observed spectra before cross-correlation, allowing for comparison with the observed line depths. The two-dimensional cross-correlation functions of the observed time-series were then subtracted from the injected two-dimensional cross-correlation functions, effectively leaving a residual signature that signifies the predicted line depth of the model. The two models are shown in Extended Data Figure\,7.

\subsection*{Determination of orbital velocity \& stellar mass}
\label{velocitydetermination}
The orbital velocity is a key observable in the application of the high-resolution cross-correlation technique \cite{2010Natur.465.1049S, brogi2012signature}. For each of the five nights, we fitted a 2-dimensional Gaussian model to the $K_{\rm p}-V_{\rm sys}$ diagram that was obtained using the template containing all 131 considered sources of line opacity at 3,000\,K. This fit provided both the best fit combination of orbital and systemic velocities for each of the nights. In the two-dimensional cross-correlation function, the radial velocity of the planetary absorption lines is expected to occur at
\vspace{-1em}
\begin{equation}
    v_{\rm RV} = v_{\rm orb} \sin{2\pi\phi} \sin{i} + v_{\rm sys,p} 
    \label{eq:2}
\end{equation}
where $\phi$ is the orbital phase of the planet and $i$ the inclination of the system. We denoted $v_{\rm orb, proj} = v_{\rm orb}\sin i = K_p$, as the projected orbital velocity as seen from Earth, equal to the radial velocity amplitude of the planet $K_p$.
\newpage
Using Kepler's 3rd law and the lever rule ($a_\ast / a_p = M_\ast / M_p$), the sum of the mass of the star and the planet is found to be
\vspace{-0.5em}
\begin{equation}
    M_\ast +  M_{\rm p} = \frac{P}{2\pi G} \left(\frac{K_\ast}{K_{\rm p}}+1\right)^3\left(\frac{v_{\rm orb, proj}}{\sin i} \right)^3 = \frac{P}{2\pi G} \left(\frac{K_\ast}{K_{\rm p}}+1\right)^3 K_p^3.
    \label{eq:kepler}
\end{equation}

Using the centre of mass of the system, the following relationship between the mass of the planet $M_p$, the mass of the star $M_\ast$, the orbital velocity $v_{\rm orb}$ (equal to the planetary radial velocity amplitude $K_p$) and the stellar radial velocity amplitude $K_\ast$ holds
\vspace{-0.5em}

\begin{equation}
    \frac{M_{\rm p}}{M_\ast}  = \frac{K_\ast}{K_{\rm p}}.
    \label{eq:binary}
\end{equation}

The combination of Eqs. \eqref{eq:kepler} and \eqref{eq:binary} results in the stellar mass given by

\begin{equation}
    M_\ast = \frac{PK_p^3}{2\pi G} \left(1 + \frac{K_\ast}{K_{\rm p}} \right)^{2}
    \label{eq:stellar_mass}
\end{equation}

Supplementary Data Table\,2 summarises the results for the projected orbital velocity, the systemic velocity and the calculated stellar mass. Based on the orbital velocity and assuming a circular orbit, we find a stellar mass of $2.08 \pm 0.14$\,$M_\odot$, which is consistent with the mass reported by Lendl et al.\,(2020) \cite{lendl2020hot} ($2.030 \pm 0.066$)
as determined via spectral synthesis modelling. \\

\noindent The radial velocity of the planet of $v_{\rm sys, p}=-27.2 \pm 0.22$\kmpers is smaller than the true systemic velocity of $-24.452 \pm 0.012$\kmpers as previously measured \cite{anderson2018wasp}, indicating the effect of atmospheric winds, blue-shifting the atmospheric absorption lines. \\

\subsection*{The stellar spectrum}
In order to determine stellar parameters of the host star we synthesised spectra using the spectral synthesis code Spectroscopy Made Easy (SME, version 580, {private communication}) \cite{sme,sme_code}, and compared them to the observed spectra. SME interpolates in a grid of one-dimensional (1D) MARCS atmosphere models \cite{marcs:08}, which are hydrostatic model atmospheres in plane parallel geometry, computed assuming LTE, chemical equilibrium, homogeneity, and conservation of the total flux (radiative plus convective, the convective flux being computed using a mixing-length recipe). This code has the advantage that it includes a flexible $\chi^2$ minimisation tool for finding the solution that fits an observed spectrum in a pre-specified spectral window. The code also includes a powerful continuum normalisation routine able to account for suppressed continuum levels as prescribed by a theoretical model, as it is evaluated against the observed spectrum. In the spectrum of a warm fast rotating star extra care is needed to normalise the spectrum, as the extreme line broadening can suppress the continuum. 
Using SME, we find the following stellar parameters, $T_{\rm eff} = 7990 \pm 90$\,K, $\log g = 3.5 \pm 0.3$, ${\rm [Fe/H]} = 0.24 \pm 0.15$, $v\sin i = 96 \pm 5$\,km\,s$^{-1}$ and $v_{\rm mic} = 2.6 \pm 0.3$\,km\,s$^{-1}$. These are consistent with values previously published\cite{lendl2020hot}. Extended Data Figure\,3 shows the spectrum of WASP-189b at the position of three out of the twenty analysed Fe lines.

\subsection*{Transmission spectroscopy of the Na doublet}
For the extraction of the planetary sodium lines in the observations of the nights of 2018-03-26, 2019-04-14 , 2019-04-25 and 2019-05-14, we followed previous work \cite{hoeijmakers2020W121, Seidel2019hot,seidel2020}. The spectra were corrected for the blaze, cosmic rays and telluric absorption lines. Telluric sodium is monitored with the detector’s fibre B on
sky and was detected in all four nights of observations. The respectively affected bins were masked during the rest of the analysis. We also masked the entire area occupied by the Doppler-shadow of sodium, which is not expected to overlap substantially with the planetary absorption due to the near-polar orbit \cite{lendl2020hot}. For robustness, the partial transits were not included in the here presented analysis. Extended Data Figure\,6 shows the transmission spectrum of WASP-189b at the wavelength of the Na D-doublet.

The possibility of false positive detections was assessed for each night via a bootstrapping method based on Redfield et al.\,(2008) \cite{redfield2008}, where each run was performed with 15,000 iterations, for further details see previous work \cite{hoeijmakers2020W121, seidel2020, seidel2020c}. The results are shown in Supplementary Data Figure\,2.

The false positive likelihood is estimated for the two nights at 0.076\% and 0.085\% respectively. The combined line depth for the sodium doublet is of $15.3 \pm 3.1$ ($\times 10^{-4}$), equivalent to 4.9$\sigma$, following the calculation of the detection level in Hoeijmakers et al.\,(2020) \cite{hoeijmakers2020W121}.

\newpage
\subsection*{Bootstrap analysis for robustness of candidate signals}
To confirm the robustness of detected species, we performed two different types of bootstrap analyses following the approach in Hoeijmakers et al.\, (2020)\cite{hoeijmakers2020W121}. The first method tests that the signal originates uniformly from in-transit exposures and that it does not appear in out-of-transit exposures. The second method tests the distribution of candidate signals caused by correlated noise in the cross-correlation functions, essentially ensuring that the detected signal is not the result of systematic noise. Detailed descriptions of the bootstrap methods can be found in Appendix\,A of Hoeijmakers et al.\,(2020) \cite{hoeijmakers2020W121}. Only if both bootstrap methods confirm the robustness, we classified the species as a detection. The bootstrap results are shown in Supplementary Data Figures\,3-10. \\

In the example of sodium (Na), we performed two different analyses, which results in two different bootstrap results, see Supplementary Data Figures\,2 and 9. Supplementary Data Figure\,9 includes the bootstrap results for sodium based on our cross-correlation analysis. Using cross-correlations only, it is not possible to extract a sodium signal, which is why sodium is classified as tentative, see Table\,\ref{tab:best_fit_gauss}. Analysing the sodium doublet following previous work, \cite{hoeijmakers2020W121, Seidel2019hot,seidel2020}, we detect sodium at a combined line depth of $15.3 \pm 3.1$ ($\times 10^{-4}$), equivalent to 4.9$\sigma$, which we do not classify as a robust detection ($5\sigma$ limit).\\

\begin{singlespace}
\noindent
{\scriptsize \textbf{Data Availability:} Raw data as well as pipeline-reduced data from which the findings that are presented in this paper are derived, are publicly available from the data archives of the European Southern Observatory (ESO) and the Telescopio Nazionale Galileo (TNG). Cross-correlation templates and models are available upon reasonable request. Pre-computed opacity functions are publicly available via \texttt{http://dace.unige.ch/opacity}. Correspondence and request for materials should be made to B.P.} \\

\noindent
{\scriptsize \textbf{Code Availability:} The computer code for performing cross correlations is publicly available at \texttt{https://github.com/hoeijmakers/tayph/}. Documentation, instructions and a data demonstration can be found at \texttt{https://tayph.readthedocs.io}.}\\

\noindent
{\scriptsize \textbf{Acknowledgements:} We acknowledge partial financial support from the PlanetS National Centre of Competence in Research (NCCR) supported by the Swiss National Science Foundation (SNSF), the European Research Council (ERC) under the European Union’s Horizon 2020 research and innovation programme (projects Four Aces, {\sc EXOKLEIN}, {\sc Spice Dune} and Exo-Atmos with grant agreement numbers 724427 (D.E., J.V.S., H.J.H.), 771620 (K.H., C.F.), 947634 (V.B) and 679633 (L.P.), respectively), UKRI Future Leaders Fellow Grant (MR/S035214/1) (H.M.C.), Spanish State Research Agency (AEI) Projects No. PID2019-107061GB-C61 (D.B.) and No. MDM-2017-0737 Unidad de Excelencia “María de Maeztu”- Centro de Astrobiología (CSIC/INTA) (D.B.), the Märta and Eric Holmberg Endowment (B.P.) and FRQNT (R.A). R. A. is a Trottier Postdoctoral Fellow and acknowledges support from the Trottier Family Foundation. The analysis presented in this work has made use of the VALD database, operated at Uppsala University, the Institute of Astronomy RAS in Moscow, and the University of Vienna; Ian Crossfields’ Astro-Python Code library and Astropy \cite{astropy:2013, astropy:2018}. C.F. acknowledges a  University of Bern International 2021 PhD Fellowship.  K.H. acknowledges a Honorary Professorship from the University of Warwick, as well as an imminent chair professorship from the Ludwig Maximilian University in Munich. B.P., H.J.H., D.K., E.S., N.W.B., B.T., C.F., M.H., B.M.M. L.P. S.G. and K.H. are part of the Mantis network.}\\

\noindent
{\scriptsize \textbf{Author Contributions Statements:} B.P. performed the data analysis (including applying computer code originally written by H.J.H.), made all of the figures except Extended Data Figures 2 and 8, co-led the scientific vision and co-led the writing of the manuscript. H.J.H. provided the computer code that was the basis and starting point for the data analysis, mentored B.P. on data analysis techniques, co-led the scientific vision and co-led the writing of the manuscript. D.K. performed radiative transfer calculations used to construct the cross correlation templates and model spectra. E.S. performed FastChem calculations and made Extended Data Figure\,8. J.V.S. investigated the fidelity of specific spectral lines, performed supporting EulerCam observations and provided the code, expertise and results to produce Supplementary Data Figures\,6 and 7. M.L. analysed the EulerCam data and made Extended Data Figure\,2. N.W.B. co-wrote the manuscript. B.T. constructed a model of the stellar spectrum and provided technical support throughout the analysis procedure. H.J.H, D.R.A. and D.B. performed HARPS observations.  K.K. performed supporting EulerCam observations.  A.G.-M. proof-read the manuscript.  S.G. provided guidance on opacities. H.C., M.H., B.M.M. and L.P. provided substantial feedback on the manuscript.  H.J.H., D.K., J.V.S., R.A., V.B., H.C., D.E., C.F., C.L., S.G., M.O. and K.H. were all co-investigators on the ESO proposal for open time in observing period 103 that led to the procurement of the data. K.H. co-led the scientific vision, co-wrote the manuscript, guided its narrative and formulation, and assisted with formatting.}\\

\noindent
{\scriptsize \textbf{Competing Interests Statement}}
{\scriptsize The authors declare no competing interests.}\\

\noindent
{\scriptsize \textbf{Acknowledgement of Version of Record}}
{\scriptsize The Version of Record of this article is published in Nature Astronomy, and is available online at \url{https://doi.org/10.1038/s41550-021-01581-z}.}

\end{singlespace}

\newpage

\subsection*{Extended Data Figures}
 \setcounter{figure}{0}
\renewcommand{\figurename}{Extended Data Figure}

\begin{figure*}[h!] 
    \centering
    \includegraphics[trim=40 15 0 50, clip, width=0.95\textwidth]{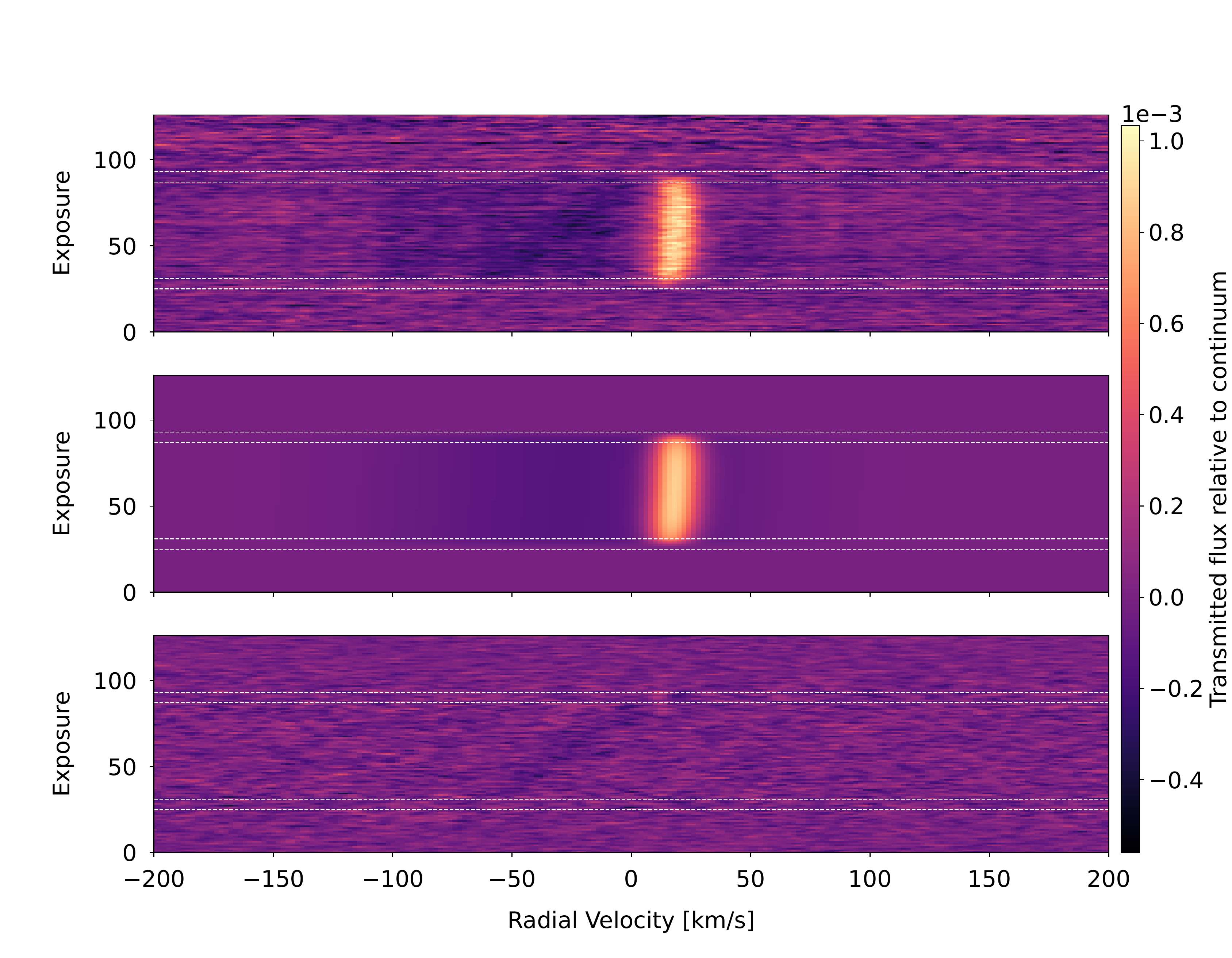}\\
    \caption{Illustration of Doppler shadow subtraction and detrending of the cross-correlation function for the time-series observed on April 14, 2019, with the Fe template at 3,000\,K (see Methods). \textbf{Top panel:} Raw two-dimensional cross-correlation function. During the transit, the Doppler shadow emerges as the positive near-vertical structure. Time of first, second, third and fourth contact as predicted using the ephemeris of Lendl et al. (2020) \cite{lendl2020hot} are indicated as dashed lines. \textbf{Middle panel:} Best-fit model of the Doppler shadow. \textbf{Bottom panel:} Residuals after subtracting the best-fit model from the raw cross-correlation function (top panel) and application of a detrending algorithm in the vertical direction. The absorption signature of the planet atmosphere is visible as the slanted feature, Doppler-shifted to the instantaneous radial velocity of the planet. The residual of the Doppler shadow at the end of the transit is masked during further analysis.}
\end{figure*}

\newpage

\begin{figure*}[h!]
    \centering
    \includegraphics[width=\textwidth]{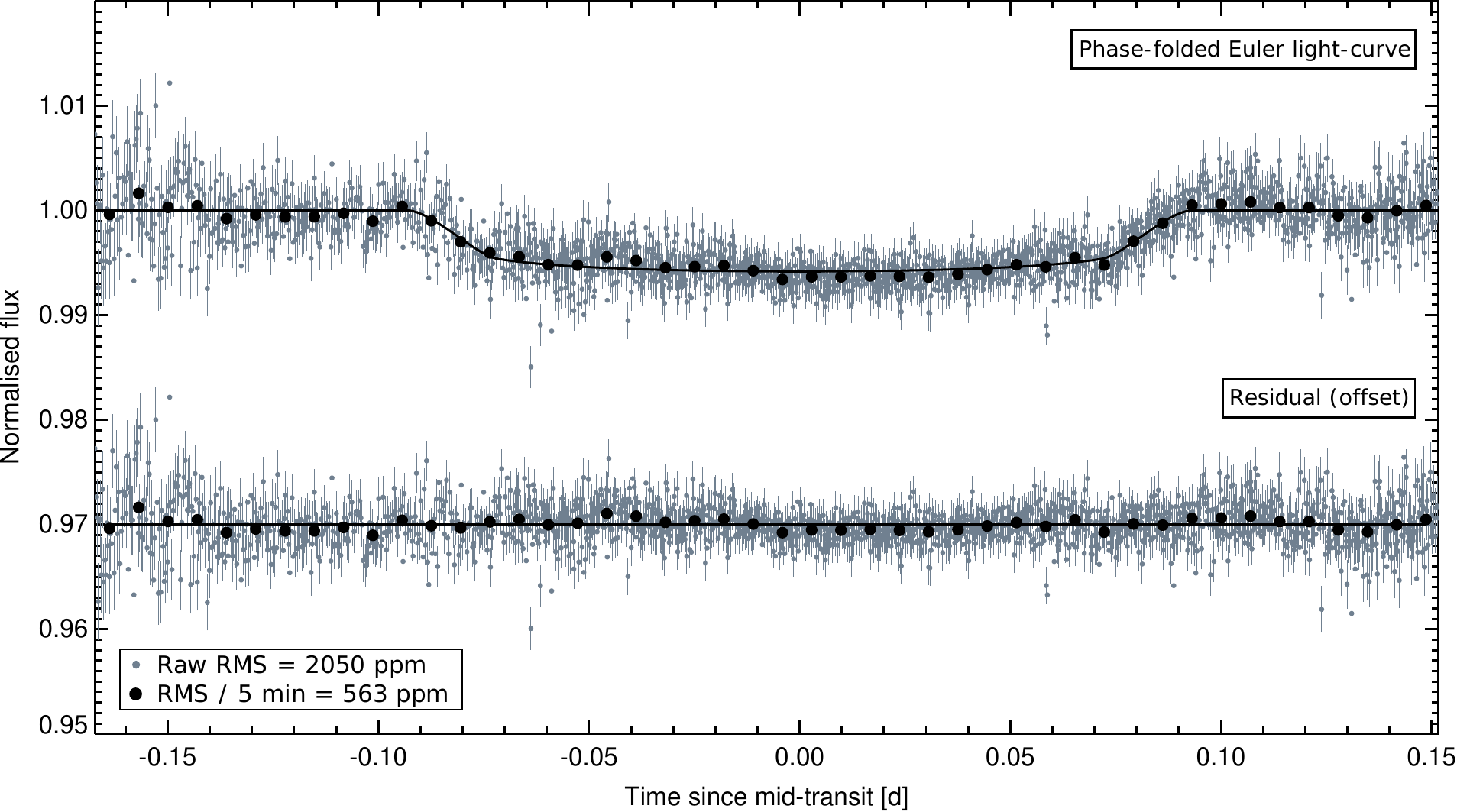}
    \caption{Phase-folded light-curve as observed with EulerCam on the nights of 14-04-2019 and 24-04-2019. No astrophysical sources of variability are detected.}
    \label{fig:euler}
\end{figure*}

\begin{figure}[h!]
    \centering
    \includegraphics[width=\textwidth]{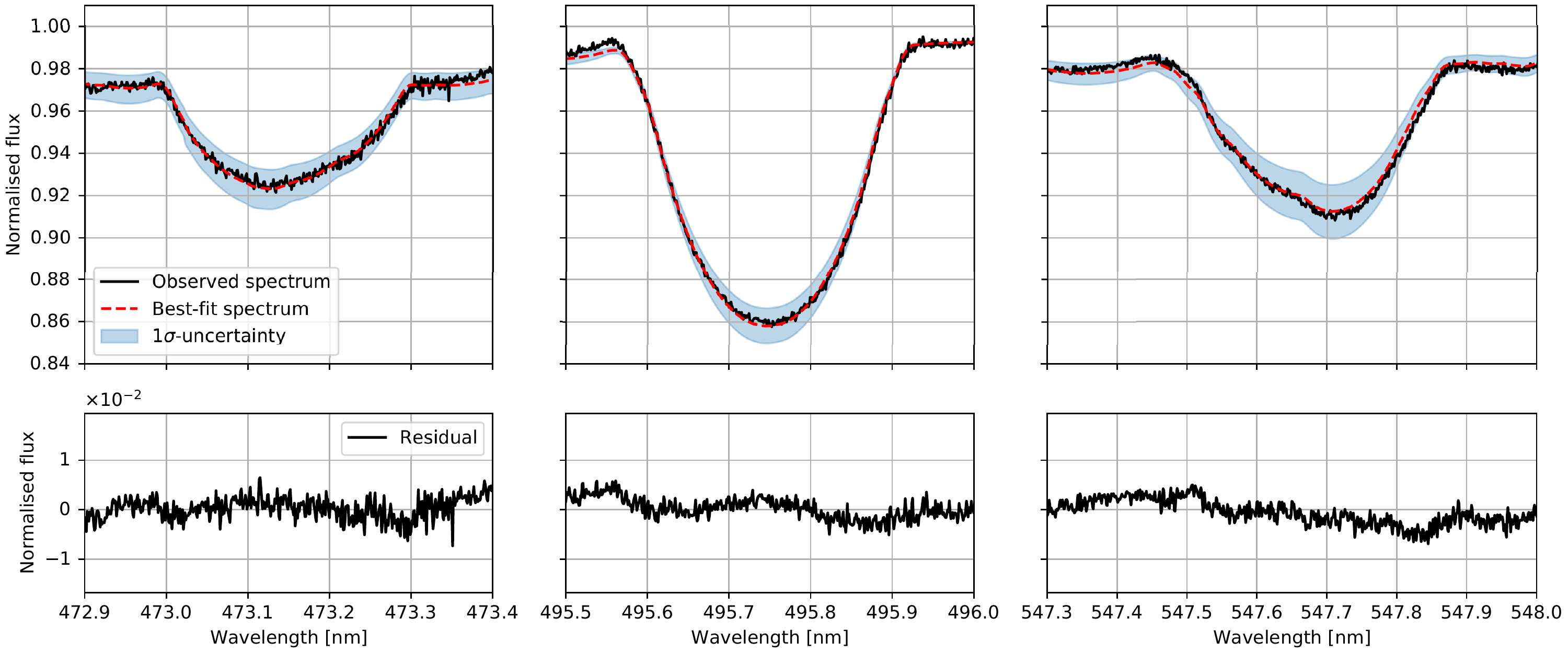}
    \caption{Three Fe lines in the spectrum of WASP-189. The black lines correspond to the observed spectrum, the red dashed lines correspond to the best fit using a metallicity of [Fe/H] = 0.24. The blue shaded regions indicate the fit with $\pm$0.15 metallicity uncertainty.}
    \label{fig:iron}
\end{figure}

\newpage

\begin{figure}[h!]
    \centering
    \includegraphics[width=0.7\textwidth]{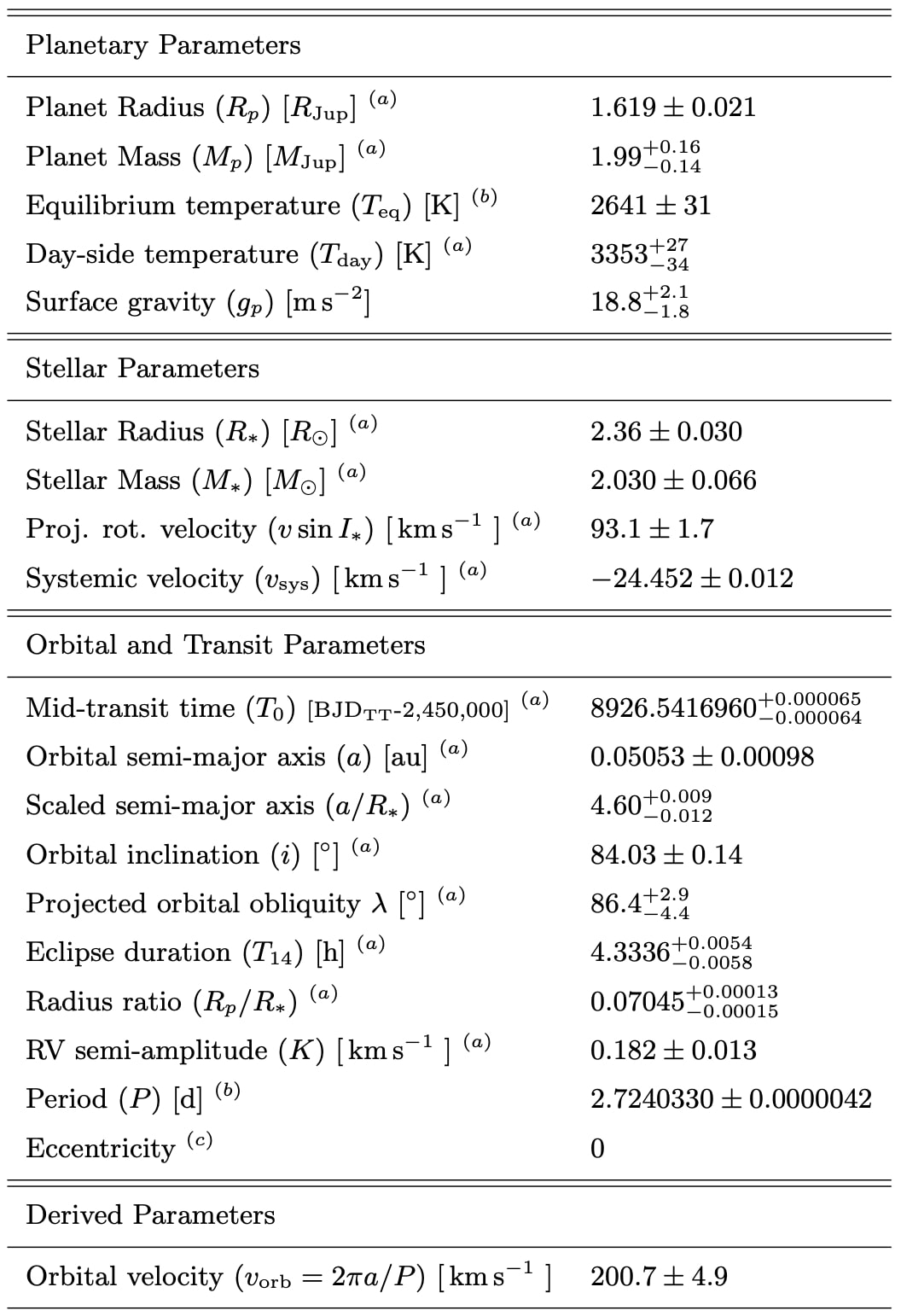}
    \caption{Summary of stellar and planetary parameters of the WASP-189 system adopted in this study. $^a$ Lendl et al. 2020\cite{lendl2020hot}, $^b$: Anderson et al. 2018\cite{anderson2018wasp} (\texttt{HARPS-MCMC}), $^c$: fixed parameter. }
\end{figure}

\begin{figure*}[h!]
    \begin{center}
    \includegraphics[width=\textwidth]{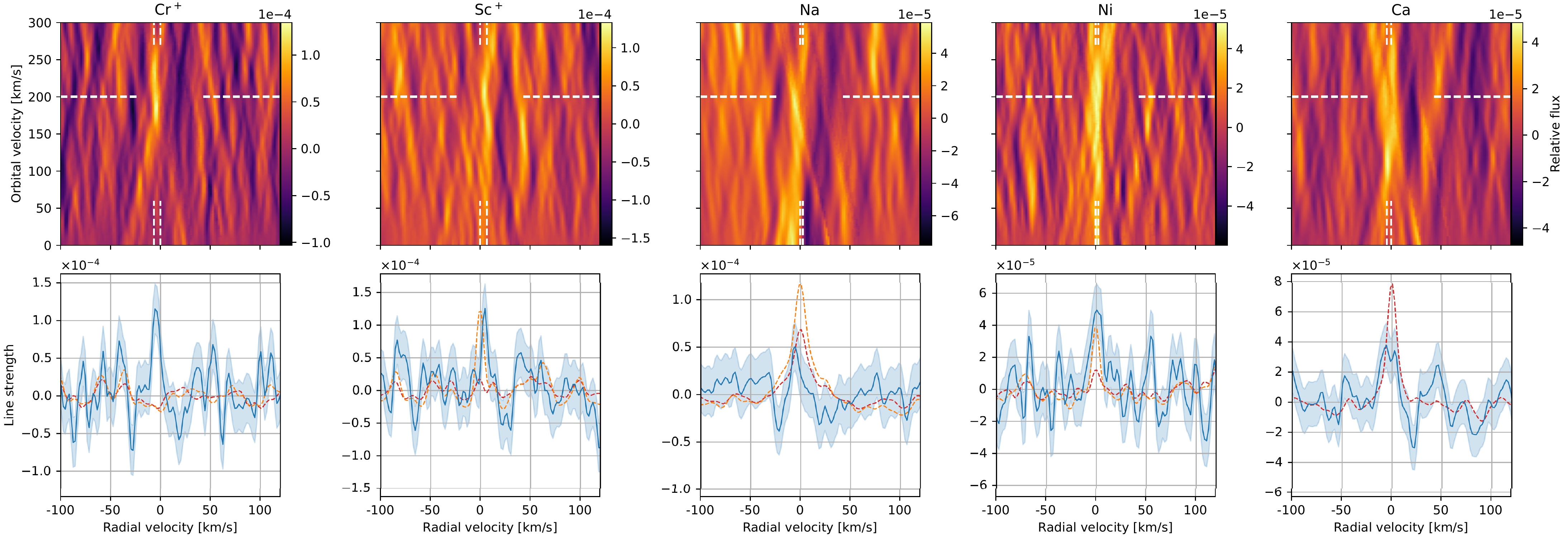}
    \end{center}
    \caption{Signals of Cr$^+$, Sc$^+$, Na, Ni and Ca classified as tentative. All of these species have previously been observed in other ultra-hot Jupiters \cite{hoeijmakers2019spectral,hoeijmakers2020W121}. The shaded region indicates the expected $1 \sigma$ uncertainty. Dashed lines show expected signal strengths, obtained by injecting and recovering the signatures of model spectra, assuming isothermal atmospheres at 2,000\,K (red) and 3,000\,K  (orange) respectively. }
\end{figure*}

\newpage

\begin{figure*}[h!]
    \begin{center}
    \includegraphics[width=0.9\textwidth]{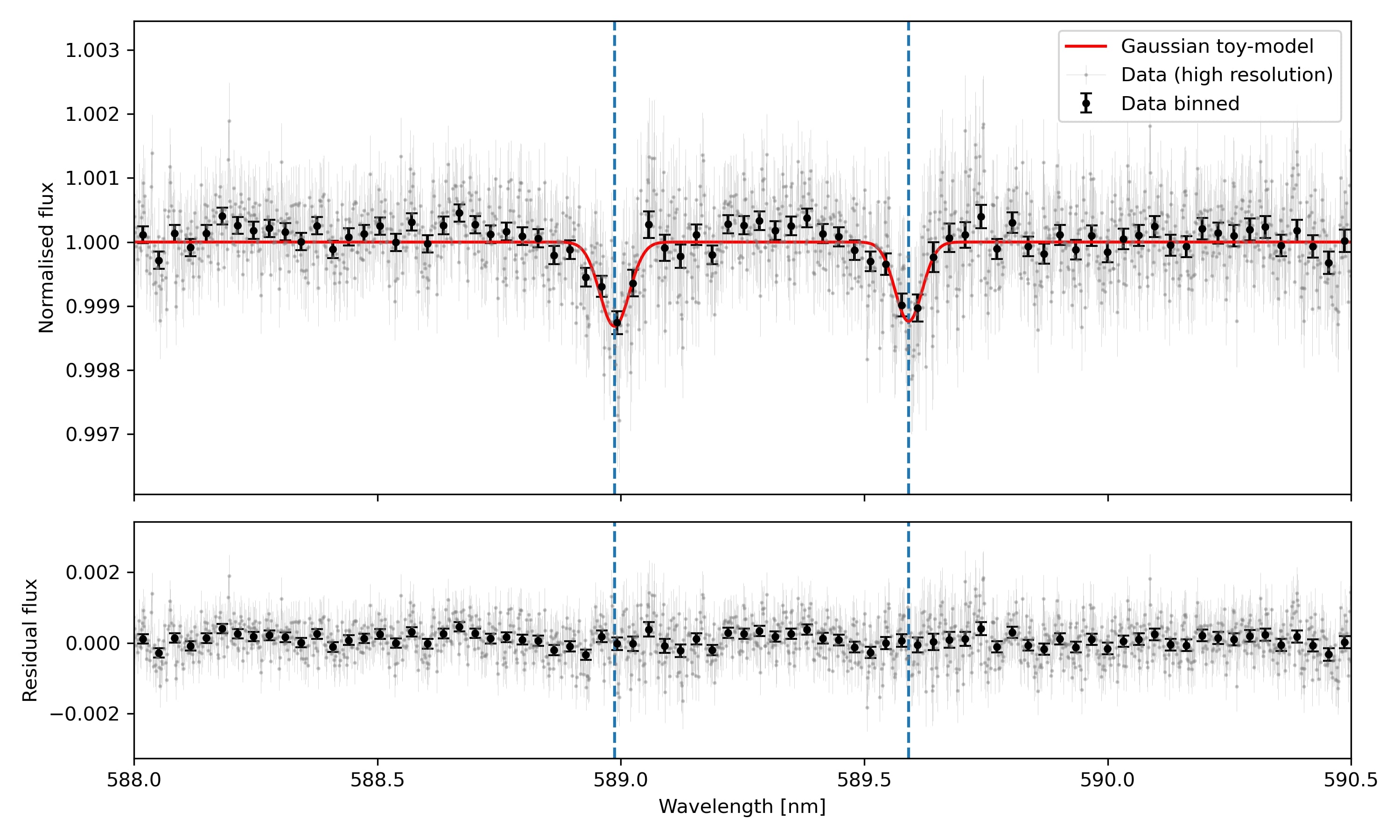}
    \end{center}
    \caption{Transmission spectrum of WASP-189b at the wavelength of the Na D-doublet. The lines are fit assuming a Gaussian line-shape, resulting in an average line depth of $15.3 \pm 3.1$ ($\times 10^{-4}$), equivalent to 4.9$\sigma$.}
\end{figure*}

\begin{figure*}[h!]
    \centering
    \includegraphics[width=\textwidth]{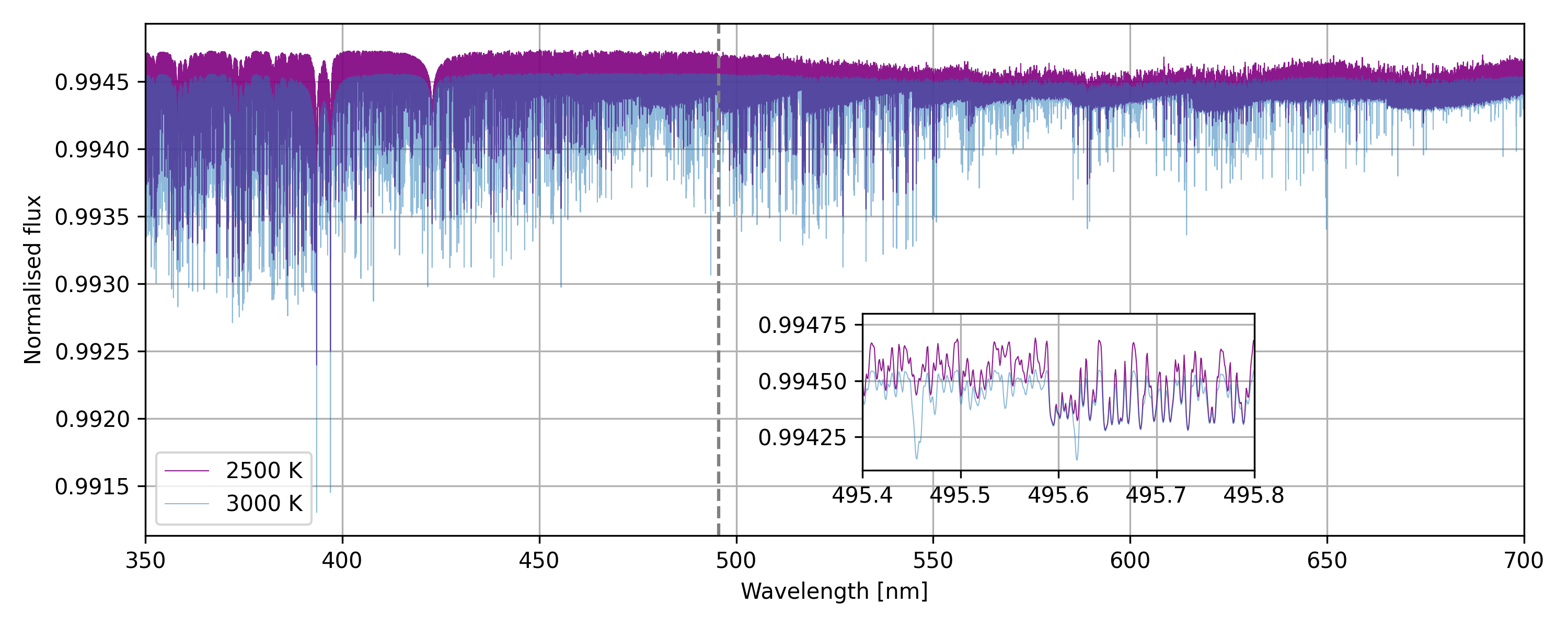}
    \caption{Two injected models of the transmission spectrum of WASP-189b at 2,500 (purple) and 3,000\,K (blue). We assumed chemical equilibrium and solar metallicity. The models are sampled at their native resolution as set by intrinsic line broadening, and not additionally broadened to match e.g. the planetary rotation or the instrumental resolving power, although such broadening terms are taken into account when injecting these templates into the data. The inset plot shows the wavelength region between 495.4 and 495.8\,nm, where a molecular band head of TiO is visible.}
\end{figure*}

\newpage

\begin{figure}[h!]
    \begin{center}
    \includegraphics[width=0.5\textwidth]{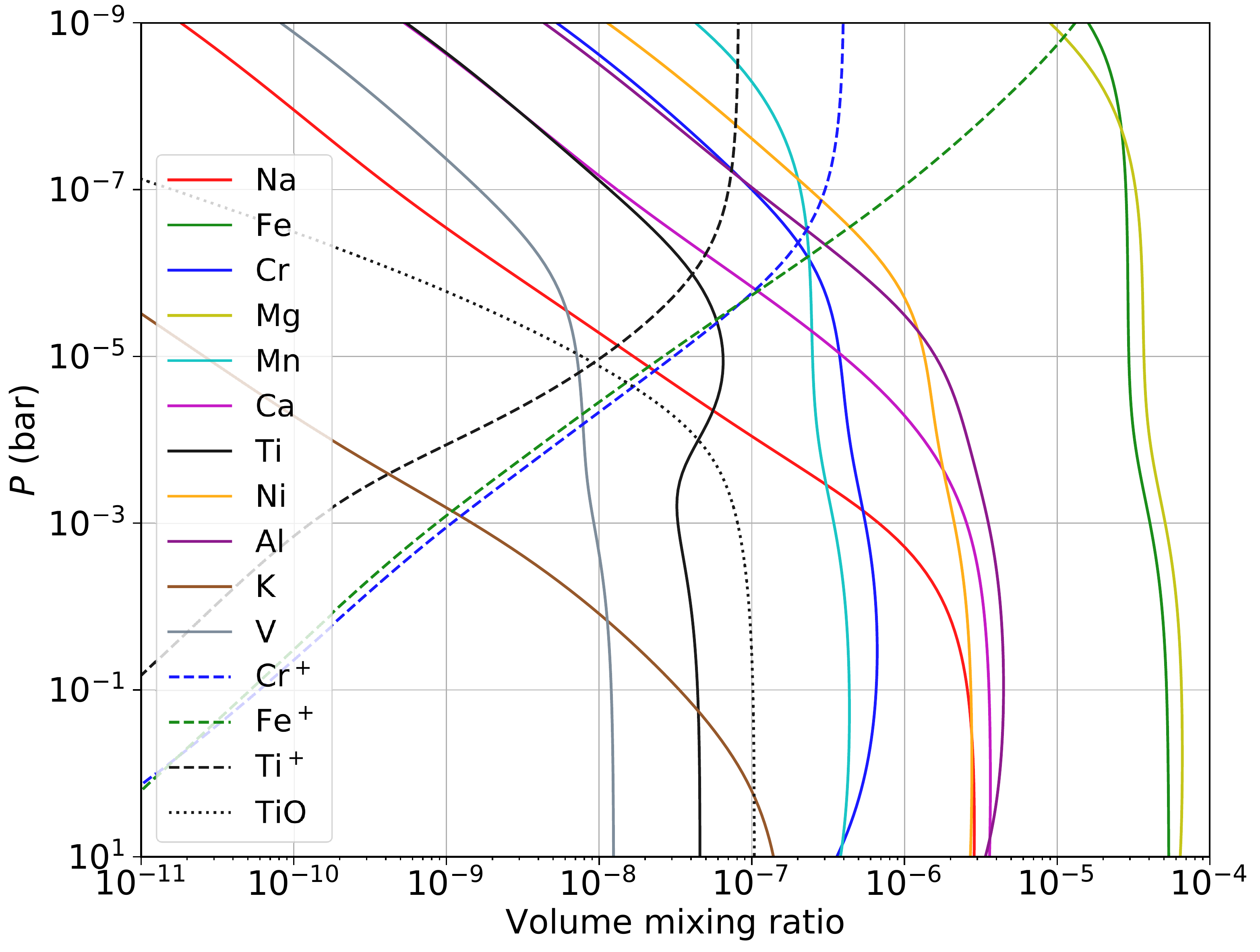}
    \end{center}
    \caption{Model of the abundances of key selected species as a function of pressure (inverse altitude) at a temperature of 2,500\,K, assuming thermo-chemical equilibrium and solar metallicity, computed with \texttt{FastChem}\cite{stock2018fastchem}. Solid lines correspond to atomic species, dashed lines to ionised species and the dotted line to TiO.}
\end{figure}

\newpage
\subsection*{Supplementary Material}
 \setcounter{figure}{0}
  \setcounter{table}{0}
 \renewcommand{\figurename}{Supplementary Data Figure}
\renewcommand{\tablename}{Supplementary Data Table}

\begin{table*}[h!]
    \begin{center}
    \begin{tabular}{cccccc}
    \hline
    \hline
         Date & $\#$Spectra$^a$ & Exp. time [s] & Airmass$^b$   & SNR @ 550 nm$^c$ & Orbital Phase $\phi$ $^d$\\
         \hline
        2018-03-26 & 39 (15/24)  & 600 & 1.97-1.11-1.52 & 144.6-213.2   & 0.99-0.096\\
        2019-04-14 & 126 (69/57) & 200 & 2.06-1.11-1.64 & 98.8-187.9    & 0.94-0.065 \\
        2019-04-25 & 107 (52/55) & 200 & 1.56-1.11-2.06 & 97.7-176.0    & 0.98-0.087 \\
        2019-05-06 & 112 (69/43) & 200 & 2.00-1.18-2.00 & 108.4-161.6   & 0.95-0.055\\ 
        2019-05-14 & 122 (68/54) & 200 & 1.94-1.11-2.27 & 90.0-136.1    & 0.93-0.047 \\
        \hline
    \end{tabular}
    \end{center}
    \captionof{table}{Overview of the observations. $^{(a)}$ In parenthesis: spectra in- and out-of-transit, respectively. $^{(b)}$ Airmass at the beginning, the peak of the altitude, and end of observation. $^{(c)}$ Minimum and maximum SNR value at 550 nm. $^{(d)}$ Orbital phase at the start and end of the observation. The transit starts at $\phi = 0.967$ and ends at $\phi = 0.033$. The observations taken in the nights of 2018-03-26 and 2019-04-25 hence only partially cover transits.}
\end{table*}

\begin{table}[h!]
    \begin{center}
    \begin{tabular}{llll}
    \hline
    \hline
         & $v_{\rm orb, proj} [{\rm km / s}]$ & $v_{\rm sys, p} [{\rm km / s}]$ & $M_\ast [M_\odot]$\\
         \hline
        2018-03-26$^\ast$ & $152.9 \pm 7.6$  & $-23.36 \pm 0.67$     & $-$ \\
        2019-04-14 & $190.8 \pm 5.3$   & $-25.67 \pm 0.25$     & $1.96 \pm 0.16$ \\
        2019-04-25 & $196 \pm 12$   & $-28.58 \pm 0.70$     & $2.13 \pm 0.40$\\
        2019-05-06 & $ 187.2 \pm 6.4$   & $-27.59 \pm 0.32$     & $1.86 \pm 0.19 $ \\
        2019-05-14 & $202.6 \pm 9.2$   & $-27.14 \pm 0.39$     & $2.35 \pm 0.32$\\
        \hline
        Mean  & $194.1 \pm 4.3$ & $ -27.2 \pm 0.22$ &  $2.08 \pm 0.14$ \\
        \hline
    \end{tabular}
    \captionof{table}{Best-fit parameters for the orbital velocity, systemic velocity and the stellar mass for all five nights of observation. $^\ast$: Due to long exposure times and only covering a half transit, this night of observations was not used to calculate the stellar mass and not taken into consideration when calculating the mean values. The best-fit orbital velocity (\vorbprecise \kmpers) is smaller than the calculated velocity using the parameters stated by Lendl et al. 2020 \cite{lendl2020hot}, $v_{\rm orb, p}$ $ = v_{\rm orb} \sin i$ $ = \frac{2 \pi a \sin i}{P}$ $ = 200.7 \pm 4.9$\kmpers, which shows that the planet's atmosphere is moving at a different speed, hence shows effects of winds. The measured systemic velocity is marginally smaller than the true systemic velocity as measured by Anderson et al. 2018 \cite{anderson2018wasp}, which can be the result of day-to-night-side winds in the atmosphere of the exoplanet.} 
    \end{center}
\end{table}

\newpage 
\begin{figure*}[h!]
    \centering
    \includegraphics[trim=40 15 0 50, clip, width=0.95\textwidth]{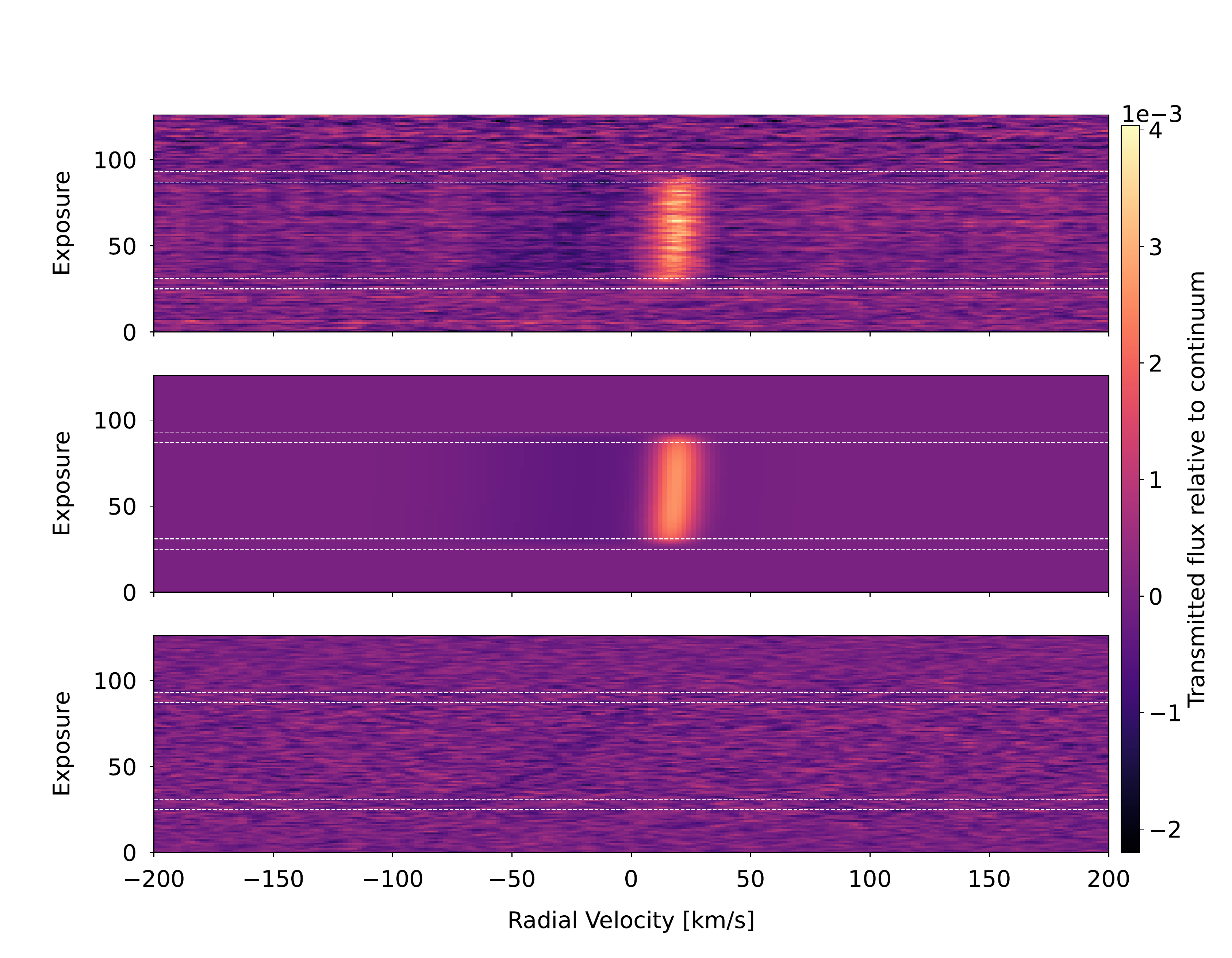}\\
    \caption{Illustration of Doppler shadow subtraction and detrending of the cross-correlation function for the time-series observed on April 14, 2019, with the Fe$^+$ template at 4,000\,K (see Methods). \textbf{Top panel:} Raw two-dimensional cross-correlation function. During the transit, the Doppler shadow emerges as the positive near-vertical structure. Time of first, second, third and fourth contact as predicted using the ephemeris of Lendl et al. (2020) \cite{lendl2020hot} are indicated as dashed lines. \textbf{Middle panel:} Best-fit model of the Doppler shadow. \textbf{Bottom panel:} Residuals after subtracting the best-fit model from the raw cross-correlation function (top panel) and application of a detrending algorithm in the vertical direction. The absorption signature of the planet atmosphere is visible as the slanted feature, Doppler-shifted to the instantaneous radial velocity of the planet. The residual of the Doppler shadow at the end of the transit is masked during further analysis.} 
\end{figure*}

\newpage

\begin{figure*}[h!]
    \begin{center}
    \includegraphics[width=\textwidth]{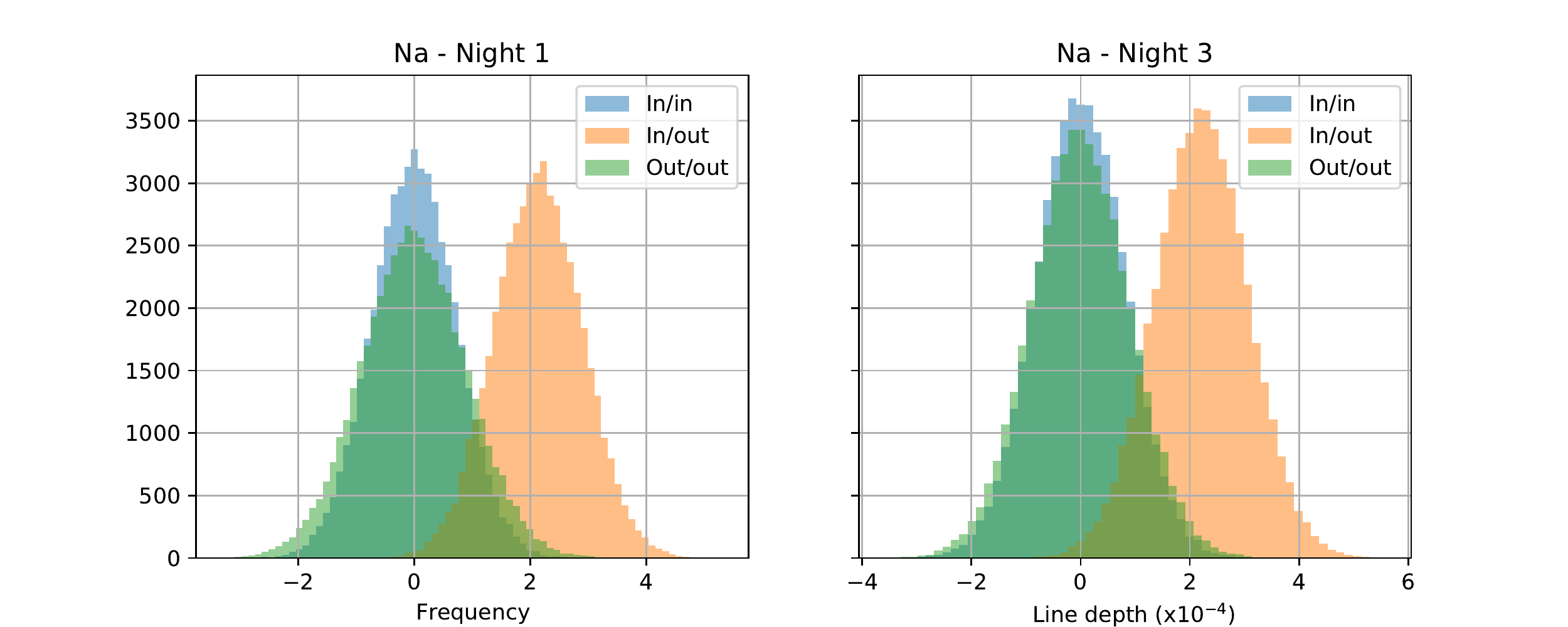}
    \end{center}
    \caption{The distributions of randomly generated instances of the one-dimensional cross-correlation function at the rest-frame velocity of the planet for the Na doublet, for each of the two nights in which the transit was fully covered. The in- and out-of-transit master functions are constructed by taking random subsets of the in-, resp. out-of-transit exposures. The measured line depth is expected to be around zero for out-out (in green) and in-in (in blue), i.e. where the randomly picked subset is normalised (subtracted) with its own master-spectrum. Only when the in-transit cross-correlation functions are normalised by the out-of-transit master spectrum (in-out, in orange), we expect the measured line amplitude to be different from zero. }
\end{figure*}

\newpage
\begin{figure*}[h!]
    \begin{center}
    \includegraphics[width=\textwidth]{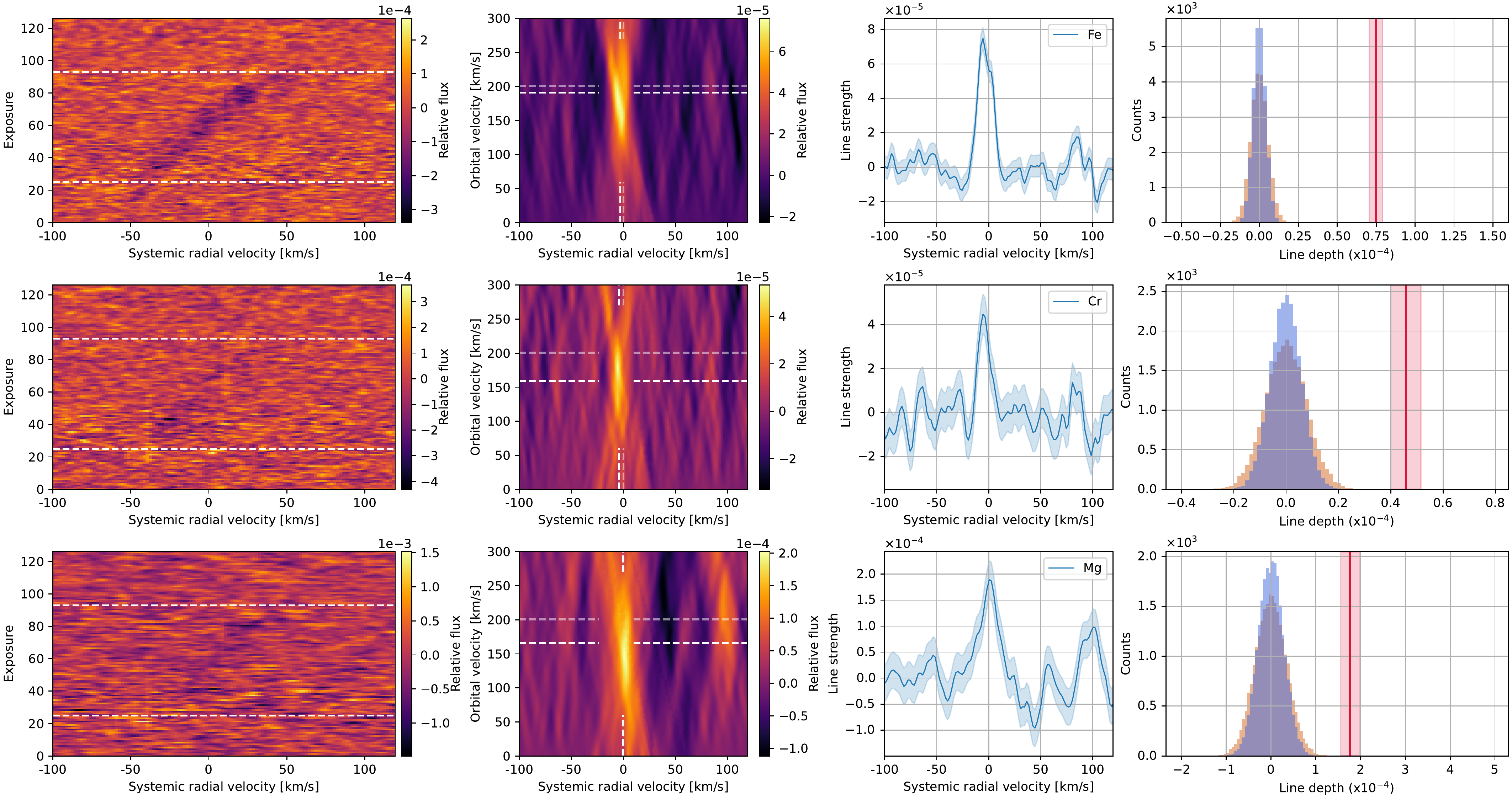}
    \end{center}
    \caption{Bootstrap results for Fe, Cr and Mg. \textbf{First and second column:} Two-dimensional exposure-velocity maps and cross-correlation functions in $K_{\rm p}-V_{\rm sys}$ space in the rest frame of the star. \textbf{Third column:} All cross-correlation functions co-added in the rest-frame of the planet shifted to the rest frame of the star, assuming orbital velocities as given in Table\,1(indicated by the strong dashed line in the second column of panels) to account for the apparent shifts due to morning-to-evening limb asymmetries. The blue-shaded area indicates the 1$\sigma$ uncertainty interval as determined through Gaussian error propagation of the expected photon noise on the individual spectra. \textbf{Forth column:} Distributions of Gaussian fits to random realisations of the one-dimensional cross-correlation function generated by stacking the time-series of cross-correlation functions with random Doppler shifts. Gaussian functions are fitted with fixed widths of 10\kmpers (in orange) and 20\kmpers (in blue). These widths correspond to typical widths observed in the real detections. These distributions are used to illustrate the robustness of the detections of each species, given the real and possibly correlated noise properties of the cross-correlation functions. The orange (10\kmpers) distributions are wider than the blue (20\kmpers) distributions. This is because systematic fluctuations that are wider are less likely to occur randomly. The red lines indicate the line depths of the detected species, and the red-shaded area the corresponding 1$\sigma$ uncertainty.}
\end{figure*}

\newpage
\begin{figure*}[h!]
    \begin{center}
    \includegraphics[width=\textwidth]{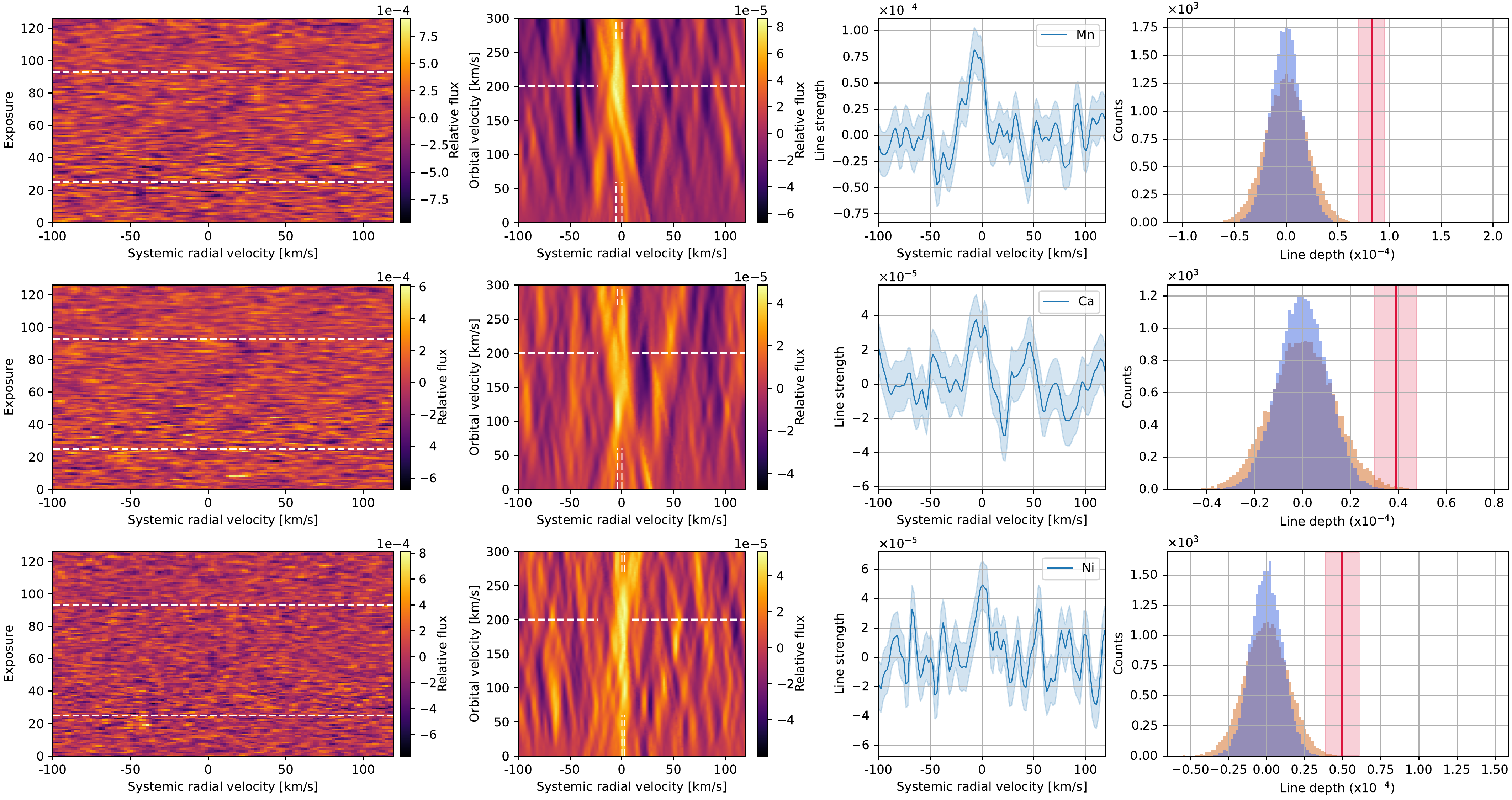}
    \end{center}
    \caption{Bootstrap results for Mn, Ca and Ni. \textbf{First and second column:} Two-dimensional exposure-velocity maps and cross-correlation functions in $K_{\rm p}-V_{\rm sys}$ space in the rest frame of the star. \textbf{Third column:} All cross-correlation functions co-added in the rest-frame of the planet shifted to the rest frame of the star, assuming orbital velocities as given in Table\,1 (indicated by the strong dashed line in the second column of panels) to account for the apparent shifts due to morning-to-evening limb asymmetries. The blue-shaded area indicates the 1$\sigma$ uncertainty interval as determined through Gaussian error propagation of the expected photon noise on the individual spectra. \textbf{Forth column:} Distributions of Gaussian fits to random realisations of the one-dimensional cross-correlation function generated by stacking the time-series of cross-correlation functions with random Doppler shifts. Gaussian functions are fitted with fixed widths of 10\kmpers (in orange) and 20\kmpers (in blue). These widths correspond to typical widths observed in the real detections. These distributions are used to illustrate the robustness of the detections of each species, given the real and possibly correlated noise properties of the cross-correlation functions. The orange (10\kmpers) distributions are wider than the blue (20\kmpers) distributions. This is because systematic fluctuations that are wider are less likely to occur randomly. The red lines indicate the line depths of the detected species, and the red-shaded area the corresponding 1$\sigma$ uncertainty.}
\end{figure*}
\newpage

\begin{figure*}[h!]
    \begin{center}
    \includegraphics[width=\textwidth]{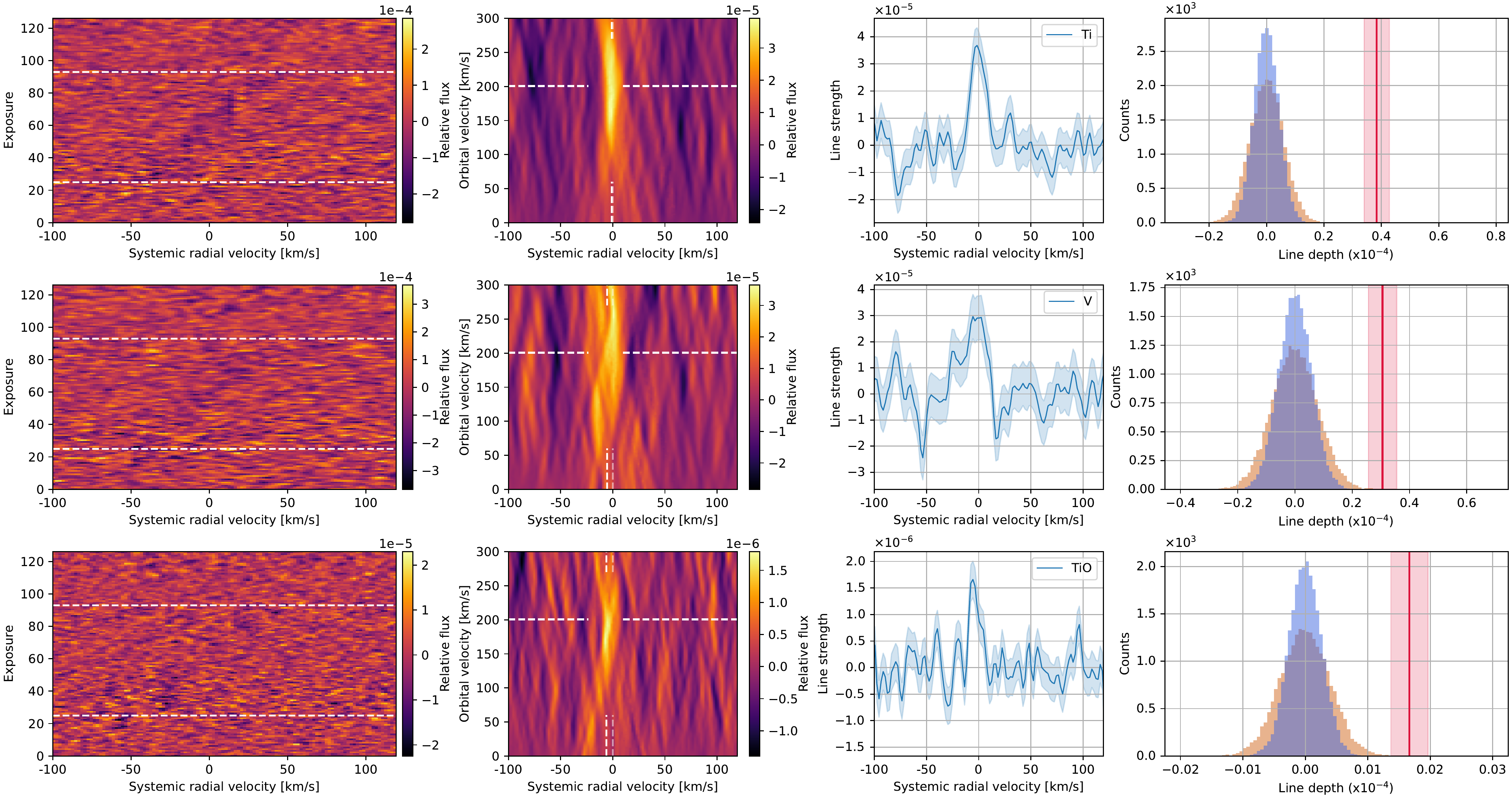}
    \end{center}
    \caption{Bootstrap results for Ti, V, and TiO. \textbf{First and second column:} Two-dimensional exposure-velocity maps and cross-correlation functions in $K_{\rm p}-V_{\rm sys}$ space in the rest frame of the star. \textbf{Third column:} All cross-correlation functions co-added in the rest-frame of the planet shifted to the rest frame of the star, assuming orbital velocities as given in Table\,1 (indicated by the strong dashed line in the second column of panels) to account for the apparent shifts due to morning-to-evening limb asymmetries. The blue-shaded area indicates the 1$\sigma$ uncertainty interval as determined through Gaussian error propagation of the expected photon noise on the individual spectra. \textbf{Forth column:} Distributions of Gaussian fits to random realisations of the one-dimensional cross-correlation function generated by stacking the time-series of cross-correlation functions with random Doppler shifts. Gaussian functions are fitted with fixed widths of 10\kmpers (in orange) and 20\kmpers (in blue). These widths correspond to typical widths observed in the real detections. These distributions are used to illustrate the robustness of the detections of each species, given the real and possibly correlated noise properties of the cross-correlation functions. The orange (10\kmpers) distributions are wider than the blue (20\kmpers) distributions. This is because systematic fluctuations that are wider are less likely to occur randomly. The red lines indicate the line depths of the detected species, and the red-shaded area the corresponding 1$\sigma$ uncertainty.}
\end{figure*}

\newpage

\begin{figure*}[h!]
    \begin{center}
    \includegraphics[width=\textwidth]{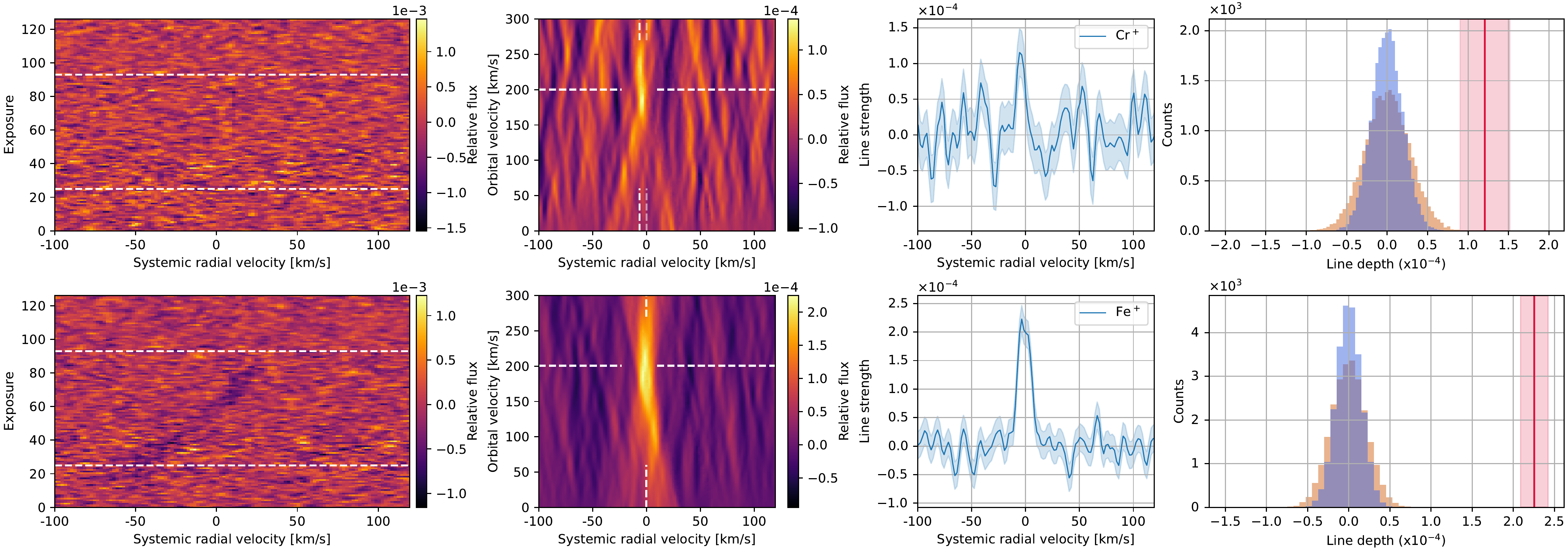}
    \includegraphics[width=\textwidth]{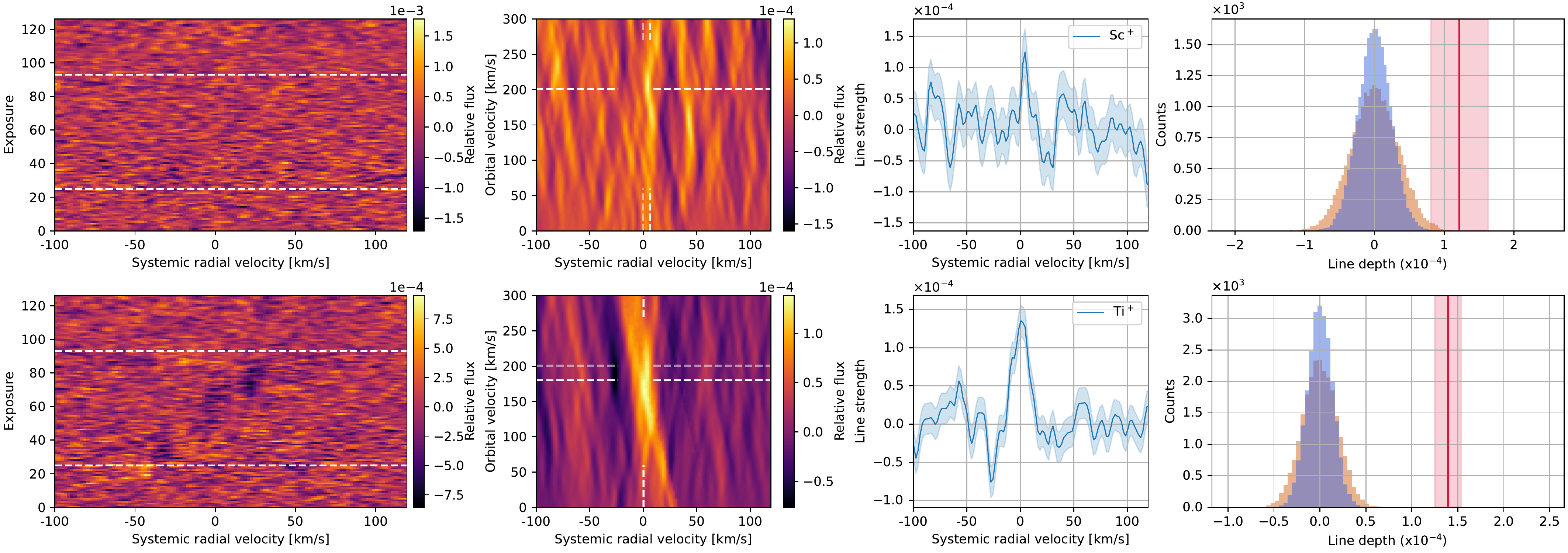}
    \end{center}
    \caption{Bootstrap results for Cr$^+$, Fe$^+$, Sc$^+$ and Ti$^+$. \textbf{First and second column:} Two-dimensional exposure-velocity maps and cross-correlation functions in $K_{\rm p}-V_{\rm sys}$ space in the rest frame of the star. \textbf{Third column:} All cross-correlation functions co-added in the rest-frame of the planet shifted to the rest frame of the star, assuming orbital velocities as given in Table\,1 (indicated by the strong dashed line in the second column of panels) to account for the apparent shifts due to morning-to-evening limb asymmetries. The blue-shaded area indicates the 1$\sigma$ uncertainty interval as determined through Gaussian error propagation of the expected photon noise on the individual spectra. \textbf{Forth column:} Distributions of Gaussian fits to random realisations of the one-dimensional cross-correlation function generated by stacking the time-series of cross-correlation functions with random Doppler shifts. Gaussian functions are fitted with fixed widths of 10\kmpers (in orange) and 20\kmpers (in blue). These widths correspond to typical widths observed in the real detections. These distributions are used to illustrate the robustness of the detections of each species, given the real and possibly correlated noise properties of the cross-correlation functions. The orange (10\kmpers) distributions are wider than the blue (20\kmpers) distributions. This is because systematic fluctuations that are wider are less likely to occur randomly. The red lines indicate the line depths of the detected species, and the red-shaded area the corresponding 1$\sigma$ uncertainty.}
\end{figure*}

\newpage
\begin{figure*}[h!]
    \begin{center}
    \includegraphics[width=\textwidth]{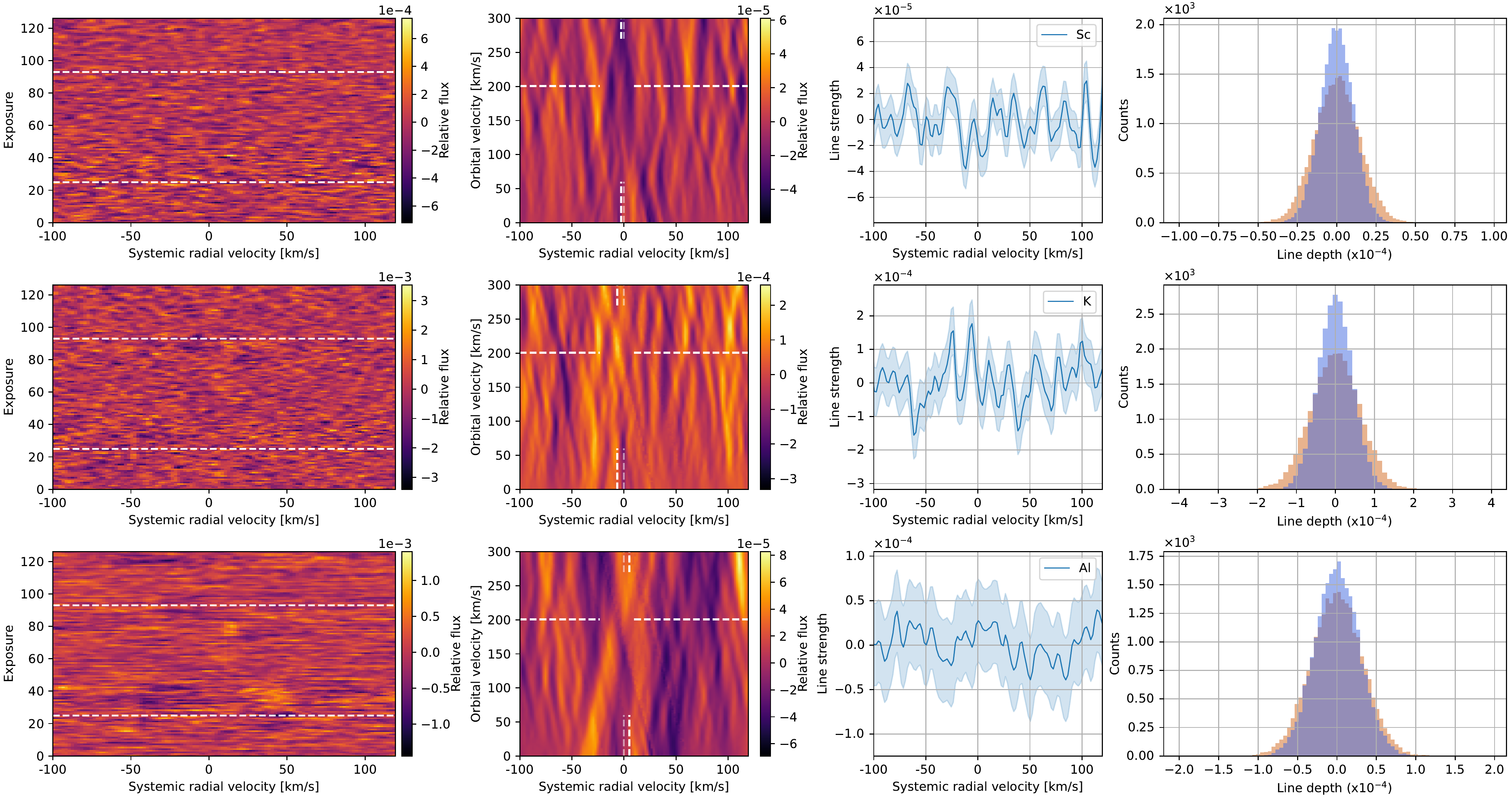}
    \end{center}
    \caption{Bootstrap results for Sc, K and Al. \textbf{First and second column:} Two-dimensional exposure-velocity maps and cross-correlation functions in $K_{\rm p}-V_{\rm sys}$ space in the rest frame of the star. \textbf{Third column:} All cross-correlation functions co-added in the rest-frame of the planet shifted to the rest frame of the star, assuming orbital velocities as given in Table\,1 (indicated by the strong dashed line in the second column of panels) to account for the apparent shifts due to morning-to-evening limb asymmetries. The blue-shaded area indicates the 1$\sigma$ uncertainty interval as determined through Gaussian error propagation of the expected photon noise on the individual spectra. \textbf{Forth column:} Distributions of Gaussian fits to random realisations of the one-dimensional cross-correlation function generated by stacking the time-series of cross-correlation functions with random Doppler shifts. Gaussian functions are fitted with fixed widths of 10\kmpers (in orange) and 20\kmpers (in blue). These widths correspond to typical widths observed in the real detections. These distributions are used to illustrate the robustness of the detections of each species, given the real and possibly correlated noise properties of the cross-correlation functions. The orange (10\kmpers) distributions are wider than the blue (20\kmpers) distributions. This is because systematic fluctuations that are wider are less likely to occur randomly.}
\end{figure*}

\newpage
\begin{figure*}[h!]
    \begin{center}
        \begin{subfigure}[b]{0.49\textwidth}
             \centering
             \includegraphics[width=\textwidth]{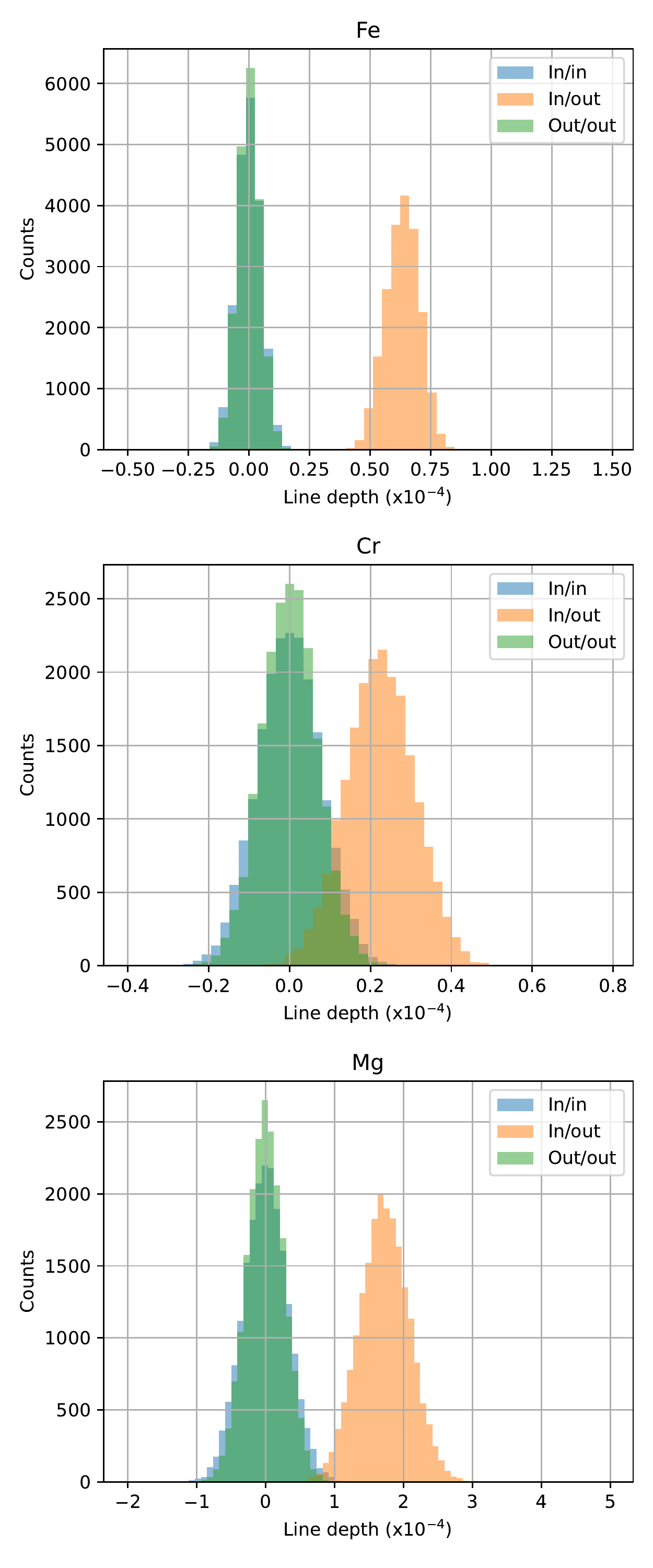}
         \end{subfigure}
         \hfill
         \begin{subfigure}[b]{0.49\textwidth}
            \centering
            \includegraphics[width=\textwidth]{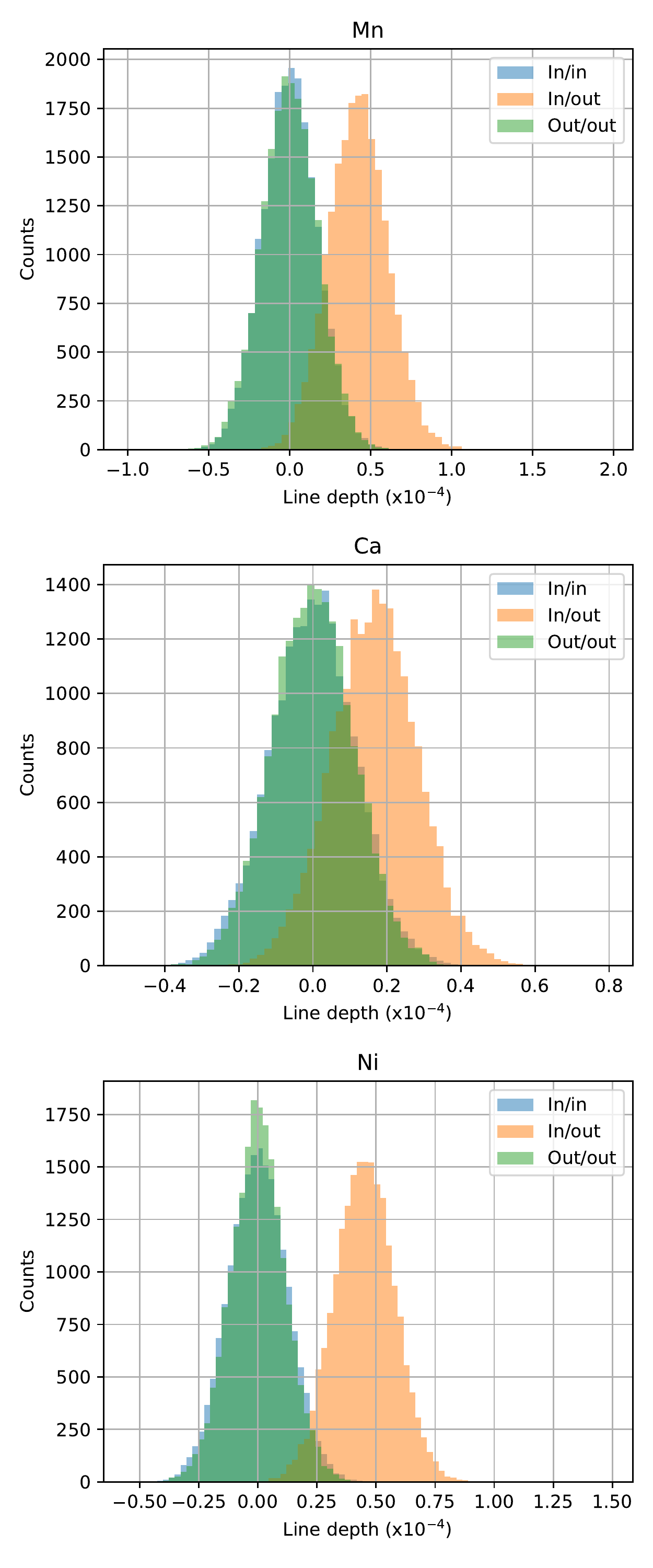}
         \end{subfigure}
    \end{center}
    \caption{The distributions of randomly generated instances of the one-dimensional cross-correlation function for Fe, Cr, Mg, Mn, Ca, and Ni at the rest-frame velocity of the planet. The in- and out-of-transit master functions are constructed by taking random subsets of the in-, resp. out-of-transit exposures. The measured line depth is expected to be around zero for out-out (in green) and in-in (in blue), i.e. where the randomly picked subset is normalised (subtracted) with its own master-spectrum. Only when the in-transit cross-correlation functions are normalised by the out-of-transit master spectrum (in-out, in orange), we expect the measured line amplitude to be different from zero.}
\end{figure*}

\newpage

\begin{figure*}[h!]
    \begin{center}
        \begin{subfigure}[b]{0.49\textwidth}
             \centering
             \includegraphics[width=\textwidth]{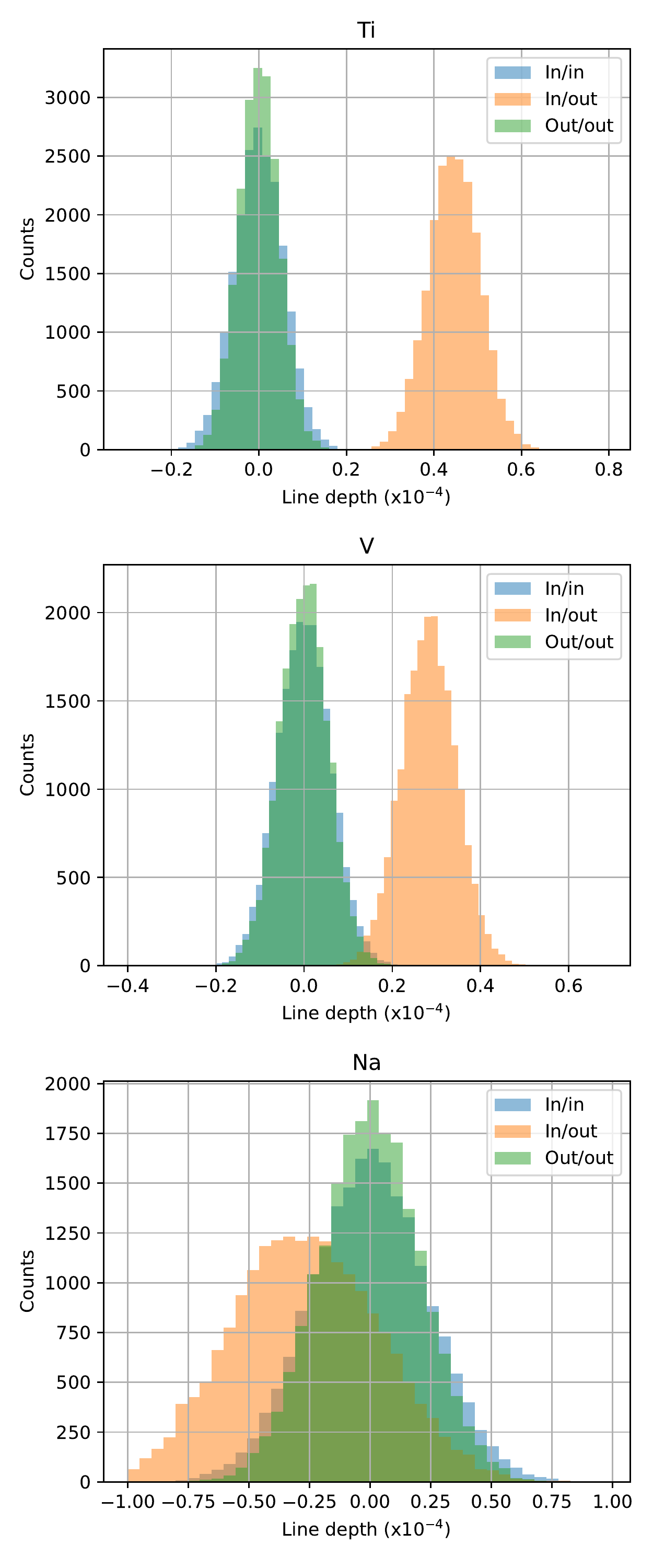}
         \end{subfigure}
         \hfill
         \begin{subfigure}[b]{0.49\textwidth}
            \centering
            \includegraphics[width=\textwidth]{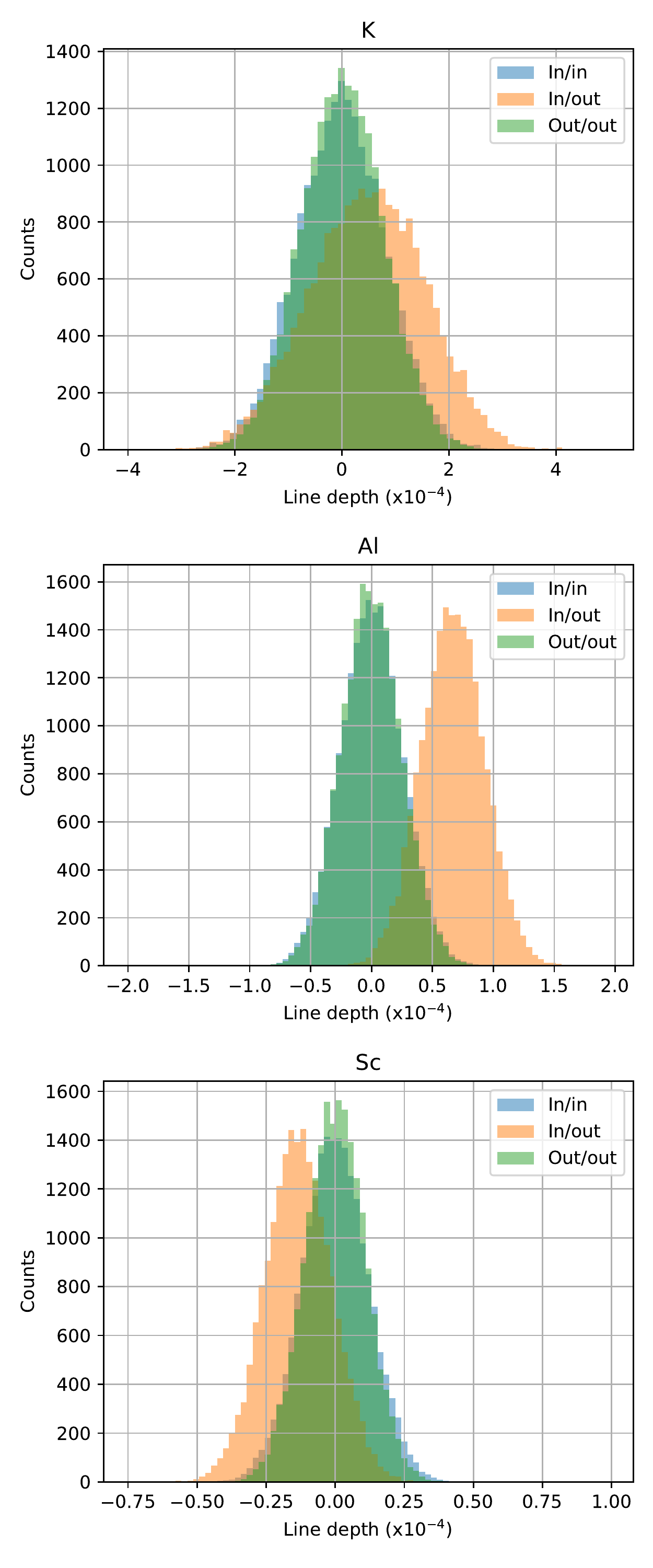}
         \end{subfigure}
    \end{center}
    \caption{The distributions of randomly generated instances of the one-dimensional cross-correlation function for Ti, V, Na, K, Al and Sc at the rest-frame velocity of the planet. The in- and out-of-transit master functions are constructed by taking random subsets of the in-, resp. out-of-transit exposures. The measured line depth is expected to be around zero for out-out (in green) and in-in (in blue), i.e. where the randomly picked subset is normalised (subtracted) with its own master-spectrum. Only when the in-transit cross-correlation functions are normalised by the out-of-transit master spectrum (in-out, in orange), we expect the measured line amplitude to be different from zero. Al seemingly presents a detection using this bootstrap method, but comparing to Supplementary Fig.\,12 shows that there is no signal present in the $K_{\rm p}-V_{\rm sys}$ diagram.}
\end{figure*}
\newpage

\begin{figure*}[h!]
    \begin{center}
        \begin{subfigure}[t]{0.49\textwidth}
            \vspace{0cm}
             \centering
             \includegraphics[width=\textwidth]{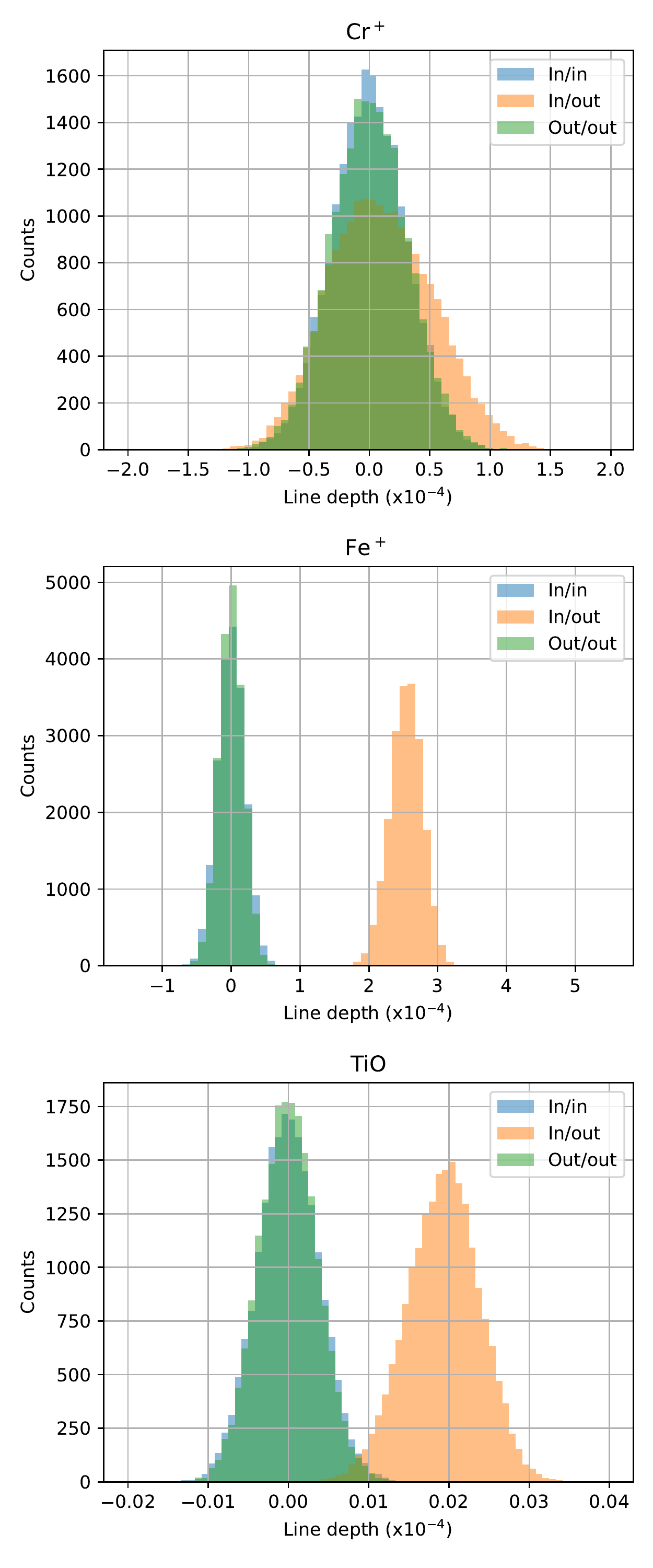}
         \end{subfigure}
         \hfill
         \begin{subfigure}[t]{0.49\textwidth}
            \vspace{0cm}
            \centering
            \includegraphics[width=\textwidth]{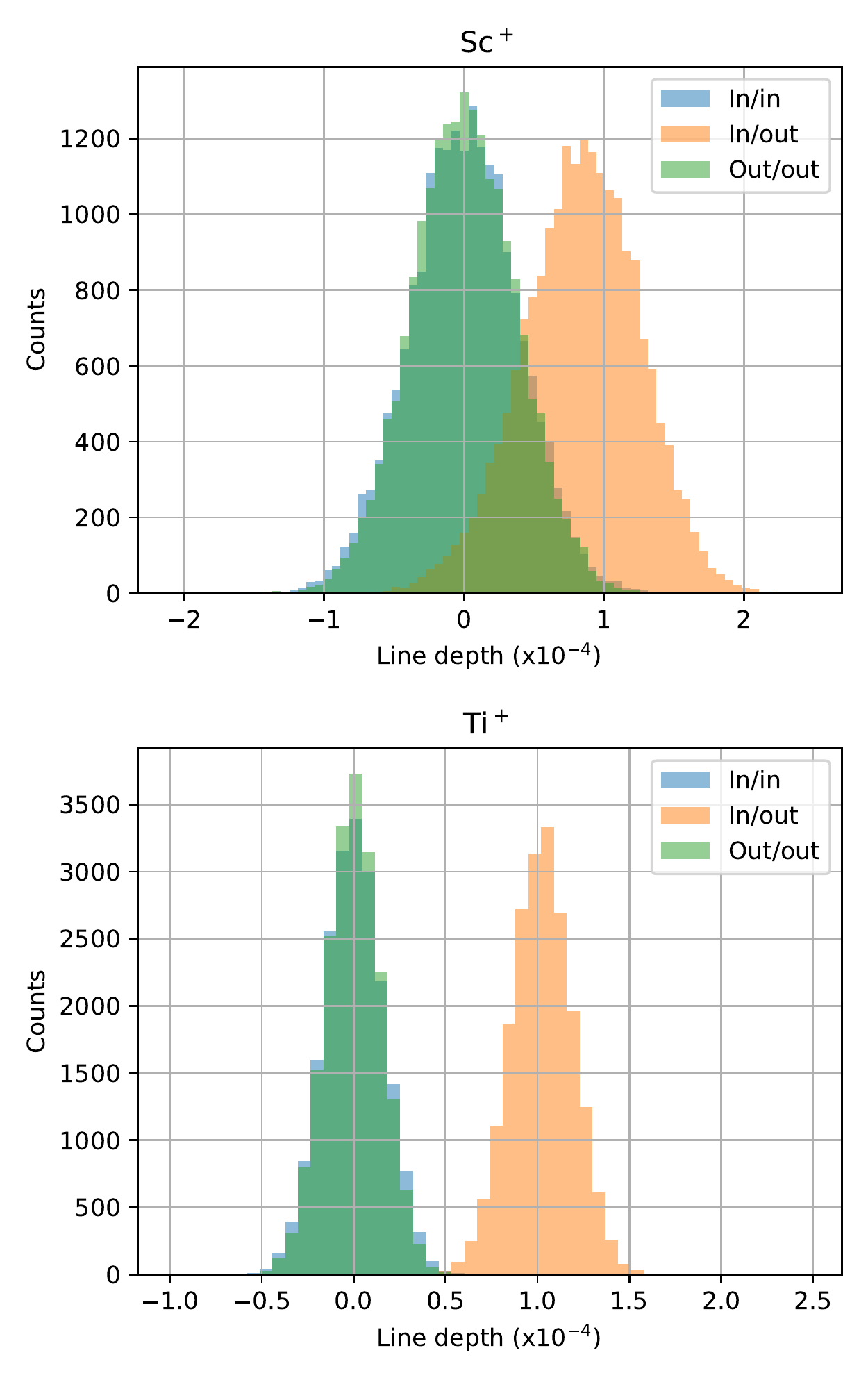}
         \end{subfigure}
    \end{center}
    \caption{The distributions of randomly generated instances of the one-dimensional cross-correlation function for Cr$^+$, Fe$^+$, TiO, Sc$^+$ and Ti$^+$ at the rest-frame velocity of the planet. The in- and out-of-transit master functions are constructed by taking random subsets of the in-, resp. out-of-transit exposures. The measured line depth is expected to be around zero for out-out (in green) and in-in (in blue), i.e. where the randomly picked subset is normalised (subtracted) with its own master-spectrum. Only when the in-transit cross-correlation functions are normalised by the out-of-transit master spectrum (in-out, in orange), we expect the measured line amplitude to be different from zero. Al seemingly presents a detection using this bootstrap method, but comparing to Supplementary Fig.\,12 shows that there is no signal present in the $K_{\rm p}-V_{\rm sys}$ diagram}
\end{figure*}
\newpage

\scriptsize
\bibliographystyle{unsrtnat}

\end{document}